\documentclass[preprint,9pt]{sigplanconf}

\usepackage{amsmath}
\usepackage[ruled,lined,linesnumbered,boxed,vlined]{algorithm2e}
\usepackage[tight,footnotesize]{subfigure}
\usepackage[squaren]{SIunits}
\usepackage{sistyle}
\usepackage[colorlinks,bookmarksopen,bookmarksnumbered,citecolor=red,urlcolor=red]{hyperref}
\usepackage{tikz}
\usetikzlibrary{positioning,shapes,shadows,arrows}
\usepackage{scalefnt}
\usepackage{subfigure}
\def\THuffman{\ensuremath{T_{\mathit{Huff}}}}
\def\TCPU{\ensuremath{T_{\mathit{CPU}}}}
\def\TGPU{\ensuremath{T_{\mathit{GPU}}}}
\def\PCPU{\ensuremath{P_{\mathit{CPU}}}}
\def\PGPU{\ensuremath{P_{\mathit{GPU}}}}
\def\TTotal{\ensuremath{T_{\mathit{total}}}}
\def\TDispatch{\ensuremath{T_{\mathit{disp}}}}

\begin{document}

\setlength{\pdfpageheight}{\paperheight}
\setlength{\pdfpagewidth}{\paperwidth}

\conferenceinfo{PMAM '14}{February 15--19, 2014, Orlando, Florida, USA} 
\copyrightyear{2014} 
\copyrightdata{978-1-nnnn-nnnn-n/yy/mm} 
\doi{nnnnnnn.nnnnnnn}




\titlebanner{}        
\preprintfooter{JPEG Decompression on Heterogeneous Multicore Architectures}   

\title{
Dynamic Partitioning-based JPEG Decompression on Heterogeneous Multicore Architectures\thanks{The definitive version of this work was published in the proceedings
of PMAM'14: Programming Models and Applications for Multicores and
Manycores, under DOI 10.1145/2560683.2560684. The
paper is available at \href{http://dl.acm.org}{dl.acm.org}} 
}
\myauthorname{
  Wasuwee~Sodsong$^1$, Jingun~Hong$^1$, Seongwook~Chung$^1$, 
  Yeongkyu~Lim$^2$,\\ Shin-Dug Kim$^1$ and Bernd~Burgstaller$^1$
}

\myauthoraffil{
  \begin{tabular}{cc}
    ${}^1$ Department of Computer Science, Yonsei University & ${}^2$ LG Electronics Inc. \\
    \{wasuwee.s, ginug, seong, sdkim\}@yonsei.ac.kr & yk.lim@lge.com\\
    bburg@cs.yonsei.ac.kr &
  \end{tabular}
}

\maketitle
\begin{abstract}
With the emergence of social networks and improvements in computational
photography, billions of JPEG images are shared and viewed on a daily
basis.  Desktops, tablets and smartphones constitute the vast majority of
hardware platforms used for displaying JPEG images.  Despite the fact that
these platforms are heterogeneous multicores, no approach exists yet
that is capable of joining forces of a system's CPU and GPU for JPEG decoding.

In this paper we introduce a novel JPEG decoding scheme for heterogeneous
architectures consisting of a CPU and an OpenCL-programmable GPU.  We employ an
offline profiling step to determine the performance of a system's CPU and GPU
with respect to JPEG decoding.  
For a given JPEG
image, our performance model uses (1)~the CPU and GPU performance
characteristics, (2)~the image entropy and (3)~the width and height of the
image to balance the JPEG decoding workload on the underlying hardware.  Our
run-time partitioning and scheduling scheme exploits task, data and pipeline
parallelism by scheduling the non-parallelizable entropy decoding task on the
CPU, whereas inverse cosine transformations (IDCTs), color conversions and
upsampling are conducted on both the CPU and the GPU.  Our kernels have been
optimized for GPU memory hierarchies.

We have implemented the proposed method in the context of the libjpeg-turbo
library, which is an industrial-strength JPEG encoding and decoding engine.
Libjpeg-turbo's hand-optimized SIMD routines for ARM and x86 constitute a
competitive yardstick for the comparison to the proposed approach.
Retro-fitting our method with libjpeg-turbo provides insights on the
software-engineering aspects of re-engineering legacy code for heterogeneous
multicores.

We have evaluated our approach for a total of 7194~JPEG images across three
high- and middle-end CPU--GPU combinations. We achieve speedups of up to 4.2x
over the SIMD-version of libjpeg-turbo, and speedups of up to 8.5x over its
sequential code.  Taking into account the non-parallelizable JPEG entropy
decoding part, our approach achieves up to 95\% of the theoretically
attainable maximal speedup, with an average of 88\%.
\end{abstract}

\category{C.1.2}{Parallel architectures}{Parallel architectures}
\category{K.6.5}{Image compression}{Image compression}

\terms
Performance, Algorithms, Design

\keywords
JPEG decoding, GPU, Parallel Programming Patterns, Pipeline-Parallelism, Data-Parallelism

\section{Introduction}
\label{sec:intro}
The JPEG format is the de facto compression standard for digital images in
a wide range of fields from medical imaging to personal digital cameras. By 
our own observation, 463 out of the 500 most popular
websites~\cite{top500sites} use JPEG images (including Google, Facebook,
Youtube and Baidu).  With the emergence of social networks and innovations in
computational photography, billions of JPEG images are shared and viewed on a
daily basis:  Facebook reported in 2010 already to store over 65~billion
photos~\cite{Beaver10findinga}, and the Instagram photo-sharing service claims
to have 45~million daily uploads~\cite{InstagramPress}.  Similar numbers can be
assumed for other photo hosting and sharing services such as Flickr and
Photobucket.

Desktops, tablets and smartphones constitute the vast majority of hardware
platforms used for viewing JPEG images. Although these platforms are nowadays
equipped with a CPU and GPU, to the best of our knowledge no approach is
available yet that is capable of incorporating both the CPU and
the GPU for JPEG decoding.
%
%
%

Libjpeg~\cite{libjpeg} is a sequential JPEG reference implementation
by the Independent JPEG Group.  To accelerate image processing, the
libjpeg-turbo~\cite{libjpegturbo} re-implementation applies SIMD instructions
with a single thread of execution on
x86 and ARM platforms.  We have observed that the SIMD-version of libjpeg-turbo
decodes
an image twice as fast as the sequential version on an Intel~i7.  Libjpeg-turbo is
widely used, e.g., with the Google Chrome and Firefox web-browsers,
WebKit~\cite{webkit}, and the Ubuntu, Fedora and openSUSE Linux distributions.
Neither libjpeg nor libjpeg-turbo are capable of utilizing a GPU.

JPEG decoding is computationally expensive, consisting of Huffman
decompression, dequantization, IDCT,
image upsampling and YCbCr to RGB color space conversion.  Among all stages, Huffman
decompression is strictly sequential, because code-words have variable
lengths and the start of a codeword in the encoded bitstream is only
known once the previous codeword has been decoded.
A sub-class of Huffman codes that provide the so-called
self-synchronization
property~\cite{huffman1988sync} are suitable for decoding multiple 
chunks of the encoded bitstream in parallel, as
proposed in~\cite{Klein03parallelhuffman}.  However, the JPEG standard
does not enforce the self-synchronization property~\cite{Wallace91thejpeg}.
In our implementation, Huffman decoding
is therefore executed sequentially on the CPU.  The remaining stages have repetitive
computations and low data dependencies, which makes them suitable to
exploit data, task and pipeline-parallelism.

A desktop GPU has several hundreds of scalar processors, offering more
parallelism than what is provided by nowaday's SIMD CPU instruction
set architectures.  GPUs offer a higher memory bandwidth than 
CPUs. 
However, a GPU core lacks complex control units and operates at a much lower
clock frequency.  The PCI bus that connects the GPU to the CPU represents a
bandwidth-bottleneck that incurs significant overhead to computations on the
GPU.  JPEG decoding on a GPU is thus not necessity faster than the SIMD-version
of libjpeg-turbo on a CPU. Nevertheless, utilizing both CPU and GPU has the
potential to achieve the highest overall performance, regardless of the
computational power of the CPU and GPU.

Consequently, this paper makes the following contributions: 
\begin{itemize}
\item
we propose a performance model based on off-line profiling that characterizes
all JPEG decoding steps for a given CPU-GPU combination
from multivariate polynomial regression. 
We identified image entropy and the image dimensions as the only 
required parameters for our performance model,
\item
we propose a dynamic partitioning scheme that automatically
distributes the workload across CPU and GPU at run-time according to 
our performance model,
\item
we optimize JPEG decoding on GPUs by employing data vectorization,
intermediate local memory and coalesced memory accesses,
\item
we boost
parallelism by utilizing a pipelined execution model that overlaps sequential
Huffman decoding with GPU computations,
\item
we report on the software
engineering aspects of refactoring the libjpeg-turbo legacy code for
heterogeneous multicores, and
\item
we present experimental results on three
representative high- and mid-end CPU-GPU architectures for the applicability
and efficiency of our approach. We achieve speedups up to 4.2x over the
SIMD-version of libjpeg-turbo, and up to 8.5x over its sequential code. 
We achieve up to 95\% of the theoretically attainable
speedup, with 88\% on average.
\end{itemize}

The remainder of this paper is organized as follows:  
Section~\ref{sec:jpegdecoding} presents background information on JPEG
decoding; libjpeg-turbo re-engineering is discussed in 
Section~\ref{sec:reengineering}.  Our OpenCL JPEG decoding kernels for GPUs
are presented in Section~\ref{sec:paralleljpeg}.
Section~\ref{sec:heterogjpeg} describes the performance model and dynamic
partitioning scheme for a system consisting of a CPU and a GPU.
Section~\ref{sec:experiment} contains the experiential
results.  We discuss the related work in Section~\ref{sec:related}
and draw our conclusions in Section~\ref{sec:conclusions}.

\section{Background: JPEG Decoding}
\label{sec:jpegdecoding}
Figure~\ref{fig:decoderpath} describes the decoding steps to produce an
uncompressed bitmap from a JPEG image. A JPEG file is structured
as a sequence of segments, including image dimensions, component subsampling,
Huffman and quantization tables and entropy-coded data.  Entropy-coded data
is the largest part of a JPEG file, and thus, has the highest contribution to
the file size.  Color in JPEG-encoded images is represented by luminance (Y),
blue chrominance (Cb) and red chrominance (Cr) component values.  Because the
human eye is
more sensitive to changes in luminance than changes in chrominance, the
spatial resolution of chrominance components are commonly compressed.  This
process is called downsampling. In 4:2:2 subsampling, the Y component is sampled
at each pixel while Cb and Cr components are sampled every two pixels in
horizontal direction.  4:4:4 sampling has the same
sample rate across all components and downsampling is not
required~\cite{poynton2002chroma}. 

\tikzstyle{djpegrect}=[rectangle, draw=black, rounded corners, fill=white, drop shadow,
        text centered, anchor=north, text=black, text width=1.9cm,minimum height=1cm]
\tikzstyle{djpegarrow}=[line width=2pt,draw, -latex]
\begin{figure}[htb]
	\begin{center}
	{\scalefont{0.9}
	\begin{tikzpicture}[node distance=0.22cm]
		\node (Huffman) [djpegrect, rectangle]
        		{Entropy Decoding (Huffman)};
		\node (AuxNode01) [text width=0.22cm, right=of Huffman] {};
		\node (Dequantization) [djpegrect, rectangle , right=of AuxNode01]
			{De-quantization};
		\node (AuxNode02) [text width=0.22cm, right=of Dequantization] {};
		\node (IDCT) [djpegrect, rectangle, right=of AuxNode02]
			{Inverse DCT};
		\node (AuxNode03) [text width=0.22cm, below=of AuxNode02] {};
		\node (Upsampling) [djpegrect, rectangle, below=of AuxNode03]
        		{Upsampling};
	    	\node (AuxNode04) [text width=0.22cm, left=of Upsampling] {};
    		\node (Color) [djpegrect, rectangle, left=of AuxNode04]
			{Color Conversion};
        
		\draw[djpegarrow] (Huffman.east) -- (Dequantization.west);  
		\draw[djpegarrow] (Dequantization.east) -- (IDCT.west); 
		\draw[djpegarrow] (IDCT.south) |- (Upsampling.east);  
		\draw[djpegarrow] (Upsampling.west) -- (Color.east); 
	\end{tikzpicture}
	}
	\caption{JPEG decoder path.}
	\label{fig:decoderpath}
	\end{center}
\end{figure}
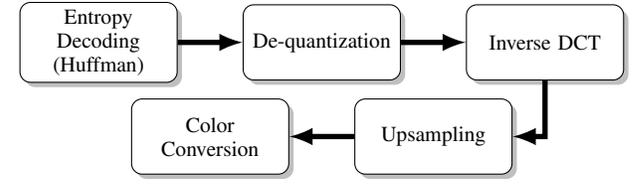

An image is divided into blocks of 8x8 pixels. JPEG decompression operates in
units of minimum coded units (MCUs), which are minimum sets of blocks from each color
component.  The MCU size for 4:4:4 subsampling is 8x8 pixels in libjpeg-turbo.
In 4:2:2 subsampling, one chrominance block is upsampled to two blocks
horizontally.  Thus, an MCU has a size of 16x8 pixels.  The decoder first
decodes entropy data, then de-quantizes it according to an image-specific
quantization table.  IDCT transforms MCUs from the frequency domain back to the
spatial domain.  The libjpeg and libjpeg-turbo libraries apply a series of 1D
IDCTs based on the AAN
algorithm~\cite{yukihiro1988fast,JPEGcompressionStandard}.  If the image
subsampling is not 4:4:4, it must be upsampled to restore the spatial
resolution of chrominance components to the original size.  Color conversion
converts the Y, Cr and Cb samples of each pixel to the RGB color space.  Apart
from Huffman decoding, all JPEG decoding steps contain few data dependencies and
a large amount of data-parallelism.  In fact, libjpeg-turbo utilizes SIMD instructions
for all stages except Huffman decoding.

\section{Re-engineering the Libjpeg-turbo Software Architecture for Heterogeneous
Multicores}
\label{sec:reengineering}

Libjpeg-turbo has been designed with a consideration of memory
resources. Both encoder and decoder operate in units of MCU rows.
The software architecture of libjpeg-turbo is illustrated in Figure~\ref{fig:arch}.
The decoder uses a 2-tier controller and buffer hierarchy to control
decoding and storage of a single MCU row in various stages.

\tikzstyle{turborect}=[rectangle, draw=black, rounded corners, fill=white, drop shadow,
        text centered, anchor=north, text=black, text width=1.5cm,minimum height=0.7cm]
\tikzstyle{turbocircle}=[circle, draw=black, fill=white, drop shadow,
        text centered, anchor=north, text=black,minimum size=0.9cm]
\tikzstyle{line}=[draw]
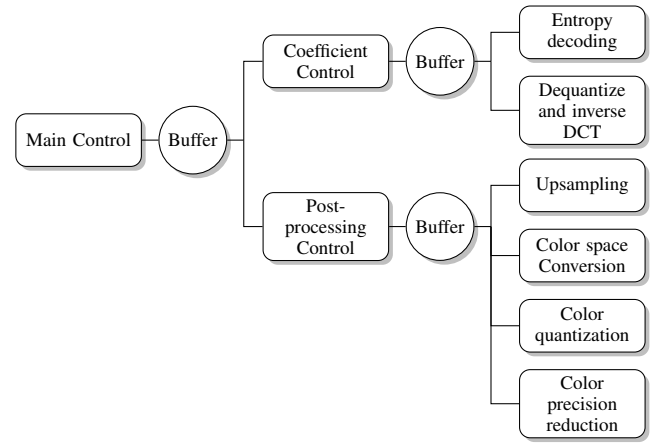
\begin{figure}[htb]
\begin{center}
{\scalefont{0.8}
\begin{tikzpicture}[node distance=0.22cm]
    \node (Main) [turborect, rectangle]
        {Main Control};
    \node (B1) [turbocircle, right=of Main]{Buffer};
    \node (AuxNode01) [text width=2cm, right=of B1] {};
    \node (Coefficient) [turborect, rectangle , above=0.6cm of AuxNode01]
        {Coefficient Control};

    \node (B2) [turbocircle, right=of Coefficient]{Buffer};
    \node (AuxNode02) [text width=0.22cm, right=of B2] {};
    \node (Postprocessing) [turborect, rectangle, below=0.6cm of AuxNode01]
        {Post-processing Control};
    \node (B3) [turbocircle, right=of Postprocessing]{Buffer};
    \node (AuxNode03) [text width=0.22cm, right=of B3] {};
    
    \node (AuxNode04) [text width=1cm, right=of AuxNode02] {};        
    \node (IDCT) [turborect, rectangle, below=0.1cm of AuxNode04,text width=1.5cm]
        {Dequantize and inverse DCT};
    \node (Huffman) [turborect, rectangle, above=of IDCT,text width=1.5cm]
        {Entropy decoding};
    \node (AuxNode05) [text width=1cm, right=of AuxNode03] {};        
    \node (Upsampling) [turborect, rectangle, above=0.1cm of AuxNode05,text width=1.5cm]
        {Upsampling};                        
    \node (Color) [turborect, rectangle, below=of Upsampling,text width=1.5cm]
        {Color space Conversion};                
    \node (Cquantization) [turborect, rectangle, below=of Color,text width=1.5cm]
        {Color quantization};                
    \node (Cprecision) [turborect, rectangle, below=of Cquantization,text width=1.5cm]
        {Color precision reduction};                                
 	\path [line]  (Main.east) -- (B1.west); 
 	\path [line]  (B1.east) -- (AuxNode01.west); 
 	\path [line]  (AuxNode01.west) |- (Coefficient.west) ;
 	\path [line]  (AuxNode01.west) |- (Postprocessing.west) ;
 	\path [line]  (Coefficient.east) -- (B2.west); 
 	\path [line]  (Postprocessing.east) -- (B3.west);  	 	
 	\path [line]  (B2.east) -- (AuxNode02.west);  	
	\path [line]  (AuxNode02.west) |- (Huffman.west) ;
 	\path [line]  (AuxNode02.west) |- (IDCT.west) ;
 	\path [line]  (B3.east) -- (AuxNode03.west);  	
	\path [line]  (AuxNode03.west) |- (Upsampling.west) ;
 	\path [line]  (AuxNode03.west) |- (Color.west) ; 	
 	\path [line]  (AuxNode03.west) |- (Cquantization.west) ; 	
 	\path [line]  (AuxNode03.west) |- (Cprecision.west) ; 	 	 	

\end{tikzpicture}
}
\caption{Libjpeg-turbo software architecture.}
\label{fig:arch}
\end{center}
\end{figure}

We identified two shortcomings which hamper parallelizing the library for
heterogeneous multicores.  First, because decoding is done in units of MCU
rows, additional, unnecessary dependencies between subsequent MCU rows are
introduced.  These dependencies limit the possible achievable parallelism in
three ways:
\begin{enumerate}
\item 
A single MCU row of an image may not contain enough data to utilize a GPU.  Most of
the computationally intensive operations, i.e., IDCT, upsampling and color
conversion, are data-parallel tasks where more data means more parallelism.
Equally worse, pipeline-parallelism between decoding steps
is impossible because of those dependencies.
\item Transferring row after row of image data from CPU to GPU is inefficient,
because initiating a transfer induces constant overhead. Transferring a large
amount of data in one transfer is thus more efficient than using several
smaller transfers. 
\item
For each MCU row, a kernel invocation on the GPU is required.
\end{enumerate}
Second, for modularity reasons, the libjpeg-turbo library has been designed
in two major parts: coefficient control and postprocessing control. 
A 2-tier buffer hierarchy abstracts away the actual decoding work,
and function pointers are used to encapsulate decoding functionality (e.g.,
integer vs.~float IDCT).  The segmented software architecture makes it very
hard to re-use JPEG decoder components, because the buffer hierarchy
permeates all components.

We re-engineered the libjpeg-turbo library under two objectives: (1)~to be
minimally invasive to the legacy code, and (2)~to support massively parallel
architectures.  To prevent expensive CPU-GPU data transfers, we introduced an
additional input and output buffer below the existing buffer hierarchy.  These
buffers are large enough to keep an image as a whole in memory.  (Note that
this fits in naturally with JPEG decoding, e.g., a web-browser will initiate
decoding once the whole image or a large part of it is available in main
memory.) The new whole-image buffers allowed us to transfer sufficiently large
chunks of the image between CPU and GPU, while providing the legacy-code on the
CPU with its ``accustomed'' row-by-row access, thereby keeping the changes to
the existing library code to a minimum. The kernel codes for the GPU were
implemented in OpenCL. The existing library code served as the starting point
for the GPU code, with all GPU-specific optimizations explained in
Section~\ref{sec:paralleljpeg}.

\section{JPEG Decoding on the GPU}
\label{sec:paralleljpeg}
After entropy (Huffman) decoding, the CPU transfers a buffer of decoded data to
the GPU.  The IDCT, upsampling and color conversion kernels are invoked
subsequently.  Our chosen buffer layout has Y blocks followed by Cb blocks
followed by Cr blocks.  The upsampling kernel does not have to read the
Y-space. This buffer layout avoids interleaving block access, and thus,
improves coalesced memory access.  At the end of color conversion, the output
image in RGB color is transferred to a designated memory location of the
whole-image output buffer (see Section~\ref{sec:reengineering}) on the CPU.
  
\subsection{Inverse Discrete Cosine Transformation (IDCT)}
\label{sub:IDCT}
The entropy-decoded data in the frequency domain is transformed back
to the spatial domain using a 2D IDCT. 
We implemented the 2D IDCT algorithm by applying a 1D IDCT to eight columns of a block (column pass) and then to eight rows of the result (row pass), as shown in Equation~\eqref{columnpass} and Equation~\eqref{rowpass} respectively. 
\begin{equation}
\label{columnpass}
f(u,y) = \sum_{v=0}^{N-1}C_vF(u,v)\cdot\cos\left(\frac{(2y+1)v\pi}{2N}\right),
\end{equation}
\begin{equation}
\label{rowpass}
f(x,y) = \sum_{u=0}^{N-1}C_uf(u,y)\cdot\cos\left(\frac{(2x+1)u\pi}{2N}\right),
\end{equation}
where 
\begin{align*}
&0\le x,y \le N-1:\thinspace \text{spatial coordinates},\\
&0\le u,v \le N-1:\thinspace \text{frequency coordinates},\\
&C_u, C_v=\frac{1}{\sqrt{2}} \text{ for } u,v = 0, \text{ otherwise } 1.
\end{align*}

We employ eight OpenCL work-items per block.  The input data is de-quantized
after being loaded from global memory.  Each work-item performs the column pass
followed by the row pass.  A work-item stores an eight-pixel column directly to
its registers such that no communication is required among work-items.  The
intermediate results from the column pass are shared among work-items within a
group to process the row pass.  Thus, local memory is the suitable choice.

Each work-item holds eight elements of 8-bit color representation at the end of
the row pass.  Copying eight times would generate an excessive overhead.  Hence, we
vectorize the elements to reduce global memory access requests.

Instructions are issued per group of work-items called a warp in NVIDIA's
terminology.  The warp size is typically 32.  Therefore, a work-group performs
IDCT on a multiple of four blocks to ensure that the number of work-items per
group is a multiple of 32.  The optimal work-group size is hardware-specific
and is determined during profiling (see Section~\ref{sec:heterogjpeg}).

\vskip+5mm
\subsection{Upsampling}
The chrominance color space with 4:2:2 subsampling is downsampled to half of the
luminance space during JPEG encoding.  The sample rates of these color spaces
must be upsampled to the original size.  Algorithm~\ref{algo:UpSampling}
describes an upsampling process that takes an 8-pixel row as an input to
generate a 16-pixel row. 

\begin{algorithm}[t]
    \SetKwInOut{Input}{Input}
    \SetKwInOut{Output}{Output}
    \SetKwData{STATE}{state}
    \SetKw{OUT}{Out}
Out[0]  = In[0] \\
Out[1]  = (In[0] * 3 + In[1] + 2) / 4 \\
Out[2]  = (In[1] * 3 + In[0] + 1) / 4 \\
Out[3]  = (In[1] * 3 + In[2] + 2) / 4 \\
Out[4]  = (In[2] * 3 + In[1] + 1) / 4 \\
Out[5]  = (In[2] * 3 + In[3] + 2) / 4 \\
Out[6]  = (In[3] * 3 + In[2] + 1) / 4 \\
Out[7]  = (In[3] * 3 + In[4] + 2) / 4 \\
Out[8]  = (In[4] * 3 + In[3] + 1) / 4 \\
Out[9]  = (In[4] * 3 + In[5] + 2) / 4 \\
Out[10] = (In[5] * 3 + In[4] + 1) / 4 \\ 
Out[11] = (In[5] * 3 + In[6] + 2) / 4 \\ 
Out[12] = (In[6] * 3 + In[5] + 1) / 4 \\ 
Out[13] = (In[6] * 3 + In[7] + 2) / 4 \\ 
Out[14] = (In[7] * 3 + In[6] + 1) / 4 \\ 
Out[15] = In[7]
    \caption{Upsampling for 4:2:2 subsampling.}
    \label{algo:UpSampling}
\end{algorithm}

We utilize 16 OpenCL work-items to perform upsampling on one block.  Two
work-items process one row of the block.  The work-item with the even ID reads
In[0] to In[4] to produce an eight-pixel row from Out[0] to Out[7], and the
work-item with the odd ID reads In[4] to In[7] to produce the successive
eight-pixel row Out[8] to Out[15]. 

The output equations have fixed patterns for odd indices and even indices.
It should be noted that all but the end pixels depend on the neighbouring
pixels of a block.  The computations of Out[0] and Out[8] happen concurrently,
and the same situation occurs with Out[7] and Out[15].  The computational
pattern of the end pixels is different  from the other pixels.  Consequently,
an if-statement is required to determine the correct equation for a specific
work-item.  An if-statement causes branch divergence if less than half of a
warp take the same branch.  We chose the work-group size such that 16
work-items take the same branch.  This access pattern was designed to favour a
merged upsampling-color conversion kernel, which is explained in
Section~\ref{KernelMerging}.

\vskip+5mm
\subsection{Color Conversion}
The final stage of the JPEG decoder converts the YCbCr color space to the RGB color
space according to Algorithm~\ref{algo:ColorSpaceConversion}.  A work-item
accesses global memory three times for its Y, Cb and Cr values to calculate R, G and
B values for one pixel. The computations for each pixel are independent
of other pixels.

\begin{algorithm}[th]
    \SetKwInOut{Input}{Input}
    \SetKwInOut{Output}{Output}
    \SetKwData{STATE}{state}
    \SetKw{OUT}{Out}
    \Input{Pixel information in YCbCr color space}
    \Output{Pixel information in RGB color space}
    R = Y + 1.402 (Cr - 128)\\
    G = Y - 0.34414 (Cb - 128) - 0.71414 (Cr - 128)\\
    B = Y + 1.772 (Cb - 128)
    \caption{Cb/Cr to RGB color space conversion.}
    \label{algo:ColorSpaceConversion}
\end{algorithm}

\begin{figure}[htb]
\begin{center}
\subfigure[Block-based pattern]
{
	\begin{tikzpicture}[scale=0.6,every node/.style={scale=0.6}]
		\foreach \i in {0,...,15} {
			\draw [very thin,black!20,text=black] (\i/2,0) -- (\i/2,4)  node [above] at (\i/2+0.25,4) {$\i$};}
		\foreach \i in {0,...,7} {
			\draw [very thin,black!20,text=black] (0,\i/2) -- (8,\i/2) node [left] at (0,4-\i/2-0.25) {$\i$};}
		\draw [very thick, black, step=4cm,xshift=0cm, yshift=0cm] (0,0) grid +(8,4);
		\draw[->,blue,thick,xshift=0.25cm,yshift=-0.25cm, rounded corners] 
			(0.0,4.0) -- (3.5,4.0) --
			(0.0,3.5) -- (3.5,3.5) --
			(0.0,3.0) -- (3.5,3.0) --
			(0.0,2.5) -- (3.5,2.5) --
			(0.0,2.0) -- (3.5,2.0) --
			(0.0,1.5) -- (3.5,1.5) --
			(0.0,1.0) -- (3.5,1.0) --
			(0.0,0.5) -- (3.5,0.5) --     
			(4.0,4.0) -- (7.5,4.0) --
			(4.0,3.5) -- (7.5,3.5) --
			(4.0,3.0) -- (7.5,3.0) --
			(4.0,2.5) -- (7.5,2.5) --
			(4.0,2.0) -- (7.5,2.0) --
			(4.0,1.5) -- (7.5,1.5) --
			(4.0,1.0) -- (7.5,1.0) --
			(4.0,0.5) -- (7.5,0.5);
	\end{tikzpicture}
	\label{fig:bufferindexing:a}
}
\quad
\subfigure[Pixel-based pattern]
{
	\begin{tikzpicture}[scale=0.6,every node/.style={scale=0.75}]
		\foreach \i in {0,...,15} {
			\draw [very thin,black!40,text=black] (\i/2,0) -- (\i/2,4)  node [above] at (\i/2+0.25,4) {$\i$};}
		\foreach \i in {0,...,7} {
			\draw [very thin,black!40,text=black] (0,\i/2) -- (8,\i/2) node [left] at (0,4-\i/2-0.25) {$\i$};}
		\draw [very thick, black, step=4cm,xshift=0cm, yshift=0cm] (0,0) grid +(8,4);
		\draw[->,blue,thick,xshift=0.25cm,yshift=-0.25cm, rounded corners] 
			(0.0,4.0) -- (7.5,4.0) --
			(0.0,3.5) -- (7.5,3.5) --
			(0.0,3.0) -- (7.5,3.0) --
			(0.0,2.5) -- (7.5,2.5) --
			(0.0,2.0) -- (7.5,2.0) --
			(0.0,1.5) -- (7.5,1.5) --
			(0.0,1.0) -- (7.5,1.0) --
			(0.0,0.5) -- (7.5,0.5);
	\end{tikzpicture}
	\label{fig:bufferindexing:b}
}
\caption{Buffer layouts of an 16x8 image \protect\subref{fig:bufferindexing:a} before color conversion and \protect\subref{fig:bufferindexing:b} after color conversion. 
The blue line indicates the access pattern. 
\label{fig:bufferindexing}}
\end{center}
\end{figure}
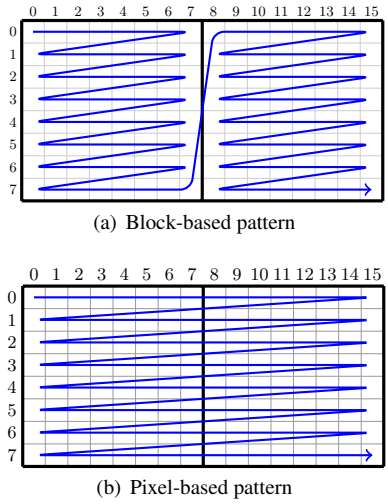

The buffers for IDCT and upsampling are arranged as a sequence of blocks, shown
in Figure~\ref{fig:bufferindexing:a}.  However, the output buffer of color
conversion is arranged as a sequence of pixels starting from the top-left pixel
of the image then traverses row-wise to the bottom-right pixel of the image as
shown in Figure~\ref{fig:bufferindexing:b}.  We devised an indexing
function that calculates the index of the next pixel in vertical
direction to be one image-width apart.

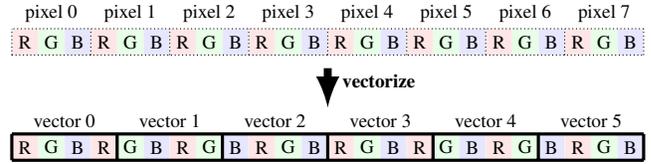
\begin{figure}[ht]
\begin{center}
\begin{tikzpicture}[scale=0.7,every node/.style={scale=0.8}]
  \foreach \j [count=\m from 0] in {0,0.5,...,3.5} 
  {
    \draw [very thin,black!20,text=black,fill=red, opacity=0.1,text opacity=1] 
          (\j*3,2) rectangle (\j*3+0.5,2.5)
          node [above] at (\j*3+0.25,2) {R};
    \draw [very thin,black!20,text=black,fill=green, opacity=0.1,text opacity=1] 
          (\j*3+0.5,2) rectangle (\j*3+1,2.5) 
          node [above] at (\j*3+0.75,2) {G};
    \draw [very thin,black!20,text=black,fill=blue, opacity=0.1,text opacity=1] 
          (\j*3+1,2) rectangle (\j*3+1.5,2.5)
          node [above] at (\j*3+1.25,2) {B};    
    \draw node [above] at (\j*3+0.75,2.5) {pixel~\m};
  }    
  
  \draw [densely dotted, black, step=1.5cm,yshift=2cm] (0,0) grid +(12,0.5);
  \draw [densely dotted, black] (12,2.5) -- (0,2.5);
  
  \foreach \j in {0,0.5,...,3.5} 
  {
    \draw [very thin,black!20,text=black,fill=red, opacity=0.1,text opacity=1] 
          (\j*3,0) rectangle (\j*3+0.5,0.5)
          node [above] at (\j*3+0.25,0) {R};
    \draw [very thin,black!20,text=black,fill=green, opacity=0.1,text opacity=1] 
          (\j*3+0.5,0) rectangle  (\j*3+1,0.5)
          node [above] at (\j*3+0.75,0) {G};
    \draw [very thin,black!20,text=black,fill=blue, opacity=0.1,text opacity=1] 
          (\j*3+1,0) rectangle (\j*3+1.5,0.5)
          node [above] at (\j*3+1.25,0) {B};   
  }
  \foreach \i in {0,...,5}
  {
      \draw node [above] at (\i*2+1,0.5) {vector~\i};
  }
  \draw [very thick, black, step=2cm] (0,0) grid +(12,0.5);
  \draw [very thick, black] (0,0.5) -- (12,0.5);  
  \coordinate (a) at (6,1.75);
  \coordinate (b) at (6,1);
  \draw[->, >=latex, black, line width=3pt]   (a) to node{} (b) node [right] at (6.1,1.5) {\textbf{vectorize}} ;
  
\end{tikzpicture}
\caption{Vectorization of interleaving RGB performed by one work-item in order to reduce global memory writes.
\label{fig:vectorizingRGB}}
\end{center}
\end{figure}

The final output of the image is represented in interleaved RGB color space.
Each R, G and B value is represented as an unsigned character.  Similar to
IDCT, vectorization in groups of four elements can be applied.  In NVIDIA's
device compute capability 2.x or higher, global memory write instructions
support 1, 2, 4, 8 or 16 bytes~\cite{OpenCLProgrammingGuide4.2}.  However, a
pixel consists of three bytes. Therefore, a work-item should perform
color conversion on a multiple of four pixels.  An eight-pixel row has 24
elements.  We group for pixels to six vectors of four elements as shown in
Figure~\ref{fig:vectorizingRGB}.  The number of transfers is thereby
reduced by a factor of four.

\vskip+5mm
\subsection{GPU Kernel Merging}
\label{KernelMerging}
Previously stored data in local memory is no longer accessible on the next
kernel invocation.  Intermediate results must be stored back to global
memory at the end of each kernel invocation, which generates
unnecessary memory traffic.
Because the computation of color conversion has no data
dependency among pixels, it can be merged with the preceding kernel to reduce
global memory accesses.

An image with 4:4:4 subsampling does not require upsampling.  The color
conversion kernel is merged with the IDCT kernel.  Color conversion requires
information from all color spaces.  Therefore, the IDCT kernel repeats the
computation three times for the three color spaces.  At the end of IDCT, a work-item
holds Y, Cb and Cr rows in its registers.  The work-item immediately performs
color conversion on the row without additional communication with
other work-items.  Although a single work-item now performs three times more
IDCT computations, the storing of intermediate results in global memory
between the IDCT and color conversion kernel invocations are avoided.

With 4:2:2 subsampling, the color conversion and upsampling kernels are combined.  We
use two OpenCL work-items to perform upsampling on a Cb and Cr row such that at
the end of upsampling, chrominance information of one row is stored in the 
registers of each work-item.  Only a row of Y space of the corresponding pixels
is loaded from global memory before starting color conversion.  Our work-group
in the merged kernel, consisting of 128 work-items, processes two groups of
four blocks.  Sixteen work-items are allocated per block, and 64 work-items
compute upsampling on the same index of different eight-pixel row segments to
avoid branch-divergence.  At the end of upsampling, this work-group produces
sixteen image blocks. 

We considered merging IDCT, upsampling and color conversion into one kernel.
Nevertheless, combining all kernels is not favourable because the number
of available registers constrains
the number of active work-groups per multiprocessor.

\subsection{Pipelined Kernel Execution}
\label{subsec:pipelinedGPU}
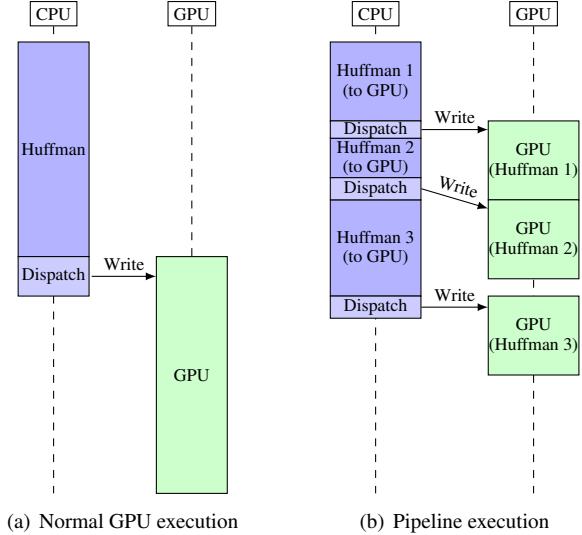
\begin{figure}[htb]
\begin{center}
\subfigure[Normal GPU execution]{
\begin{tikzpicture}[
  every node/.style={font=\normalsize,
  minimum height=0.30cm,minimum width=0.15cm},
  box/.style={minimum height=0.30cm,minimum width=1.25cm},
  timelabel/.style={font=\normalsize,midway, above, sloped},
  scale=0.75,every node/.style={scale=0.75},
  ]

\node [matrix, ampersand replacement=\&, very thin,column sep=0.15cm,row sep=0.30cm] (matrix) at (0,0) {
  \node(0,0) (duration) {}; \& \node(0,0) (CPU) {}; 	\& \& \& \& \& \&
  \node(0,0) (GPU) {}; \&  \node(0,0) (GPU time) {}; \\  
  \node(0,0) (CPU time 0) {}; \& \node(0,0) (CPU 0) [box] {};	 \& \& \& \& \& \&
  \node(0,0) (GPU 0) [box] {}; \&  \node(0,0) (GPU time 0) {}; \\
  \node(0,0) (CPU time 1) {}; \&  \node(0,0) (CPU 1) [box] {};	\& \& \& \& \& \&
  \node(0,0) (GPU 1) [box] {}; \&  \node(0,0) (GPU time 1) {}; \\
  \node(0,0) (CPU time 2) {}; \&  \node(0,0) (CPU 2) [box] {};  \& \& \& \& \& \&
  \node(0,0) (GPU 2) [box] {}; \&  \node(0,0) (GPU time 2) {};  \\
  \node(0,0) (CPU time 3) {}; \&  \node(0,0) (CPU 3) [box] {};	\& \& \& \& \& \&
  \node(0,0) (GPU 3) [box] {}; \&  \node(0,0) (GPU time 3) {};  \\
  \node(0,0) (CPU time 4) {}; \&  \node(0,0) (CPU 4) [box] {}; 	\& \& \& \& \& \&
  \node(0,0) (GPU 4) [box] {}; \&  \node(0,0) (GPU time 4) {};  \\
  \node(0,0) (CPU time 5) {}; \&  \node(0,0) (CPU 5) [box] {}; 	\& \& \& \& \& \&
  \node(0,0) (GPU 5) [box] {}; \&  \node(0,0) (GPU time 5) {};  \\
  \node(0,0) (CPU time 6) {}; \&  \node(0,0) (CPU 6) [box] {};	\& \& \& \& \& \&
  \node(0,0) (GPU 6) [box] {}; \&   \node(0,0) (GPU time 6) {};  \\
  \node(0,0) (CPU time 7) {}; \& \node(0,0) (CPU 7) [box] {};	 \& \& \& \& \& \&
  \node(0,0) (GPU 7) [box] {}; \&  \node(0,0) (GPU time 7) {}; \\
  \node(0,0) (CPU time 8) {}; \&  \node(0,0) (CPU 8) [box] {};	\& \& \& \& \& \&
  \node(0,0) (GPU 8) [box] {}; \&  \node(0,0) (GPU time 8) {}; \\
  \node(0,0) (CPU time 9) {}; \&  \node(0,0) (CPU 9) [box] {};  \& \& \& \& \& \&
  \node(0,0) (GPU 9) [box] {}; \&  \node(0,0) (GPU time 9) {};  \\
  \node(0,0) (CPU time 10) {}; \&  \node(0,0) (CPU 10) [box] {};	\& \& \& \& \& \&
  \node(0,0) (GPU 10) [box] {}; \&  \node(0,0) (GPU time 10) {};  \\
  \node(0,0) (CPU time 11) {}; \&  \node(0,0) (CPU 11) [box] {}; 	\& \& \& \& \& \&
  \node(0,0) (GPU 11) [box] {}; \&  \node(0,0) (GPU time 11) {};  \\
};

\fill 
	(CPU) node[draw,fill=white] (t1) {CPU}
	(GPU) node[draw,fill=white] (t2) {GPU};

\draw [dashed] (t1.south) -- (CPU 11.south);
\draw [dashed] (t2.south) -- (GPU 11.south);

\filldraw[fill=blue!30]
  (CPU 0.north west) rectangle (CPU 5.south east) node (huffman) [midway,align=center] {Huffman};
\filldraw[fill=blue!20]
  (CPU 5.south west) rectangle (CPU 6.south east) node (dispatch) [midway,align=center] {Dispatch};
\filldraw[fill=green!20]
  (GPU 5.south west) rectangle (GPU 11.south east) node (opencl) [midway,align=center] {GPU};
%

\draw [-latex] (dispatch.east) -- (dispatch.east -| GPU 4.west) node [timelabel] (TextNode) {Write};

\end{tikzpicture}
\label{fig:timelineGPU:a}}
\quad
\subfigure[Pipeline execution]{
\begin{tikzpicture}[
  every node/.style={font=\normalsize,
  minimum height=0.30cm,minimum width=0.15cm},
  box/.style={minimum height=0.30cm,minimum width=1.6cm},
  timelabel/.style={font=\normalsize,midway, above, sloped},
  scale=0.75,every node/.style={scale=0.75},
  ]

\node [matrix, ampersand replacement=\&, very thin,column sep=0.15cm,row sep=0.30cm] (matrix) at (0,0) {
  \node(0,0) (duration) {};	 	\& \node(0,0) (CPU) {}; 			\& \& \& \& \& \&  \node(0,0) (GPU) {}; 			\&  \node(0,0) (GPU time) {}; \\  
  \node(0,0) (CPU time 0) {}; 	\& \node(0,0) (CPU 0) [box] {};	\& \& \& \& \& \&  \node(0,0) (GPU 0) [box] {}; 	\&  \node(0,0) (GPU time 0) {}; \\
  \node(0,0) (CPU time 1) {}; 	\& \node(0,0) (CPU 1) [box] {};	\& \& \& \& \& \&  \node(0,0) (GPU 1) [box] {}; 	\&  \node(0,0) (GPU time 1) {}; \\
  \node(0,0) (CPU time 2) {}; 	\& \node(0,0) (CPU 2) [box] {}; 	\& \& \& \& \& \&  \node(0,0) (GPU 2) [box] {}; 	\&  \node(0,0) (GPU time 2) {}; \\
  \node(0,0) (CPU time 3) {}; 	\& \node(0,0) (CPU 3) [box] {};	\& \& \& \& \& \&  \node(0,0) (GPU 3) [box] {}; 	\&  \node(0,0) (GPU time 3) {}; \\
  \node(0,0) (CPU time 4) {}; 	\& \node(0,0) (CPU 4) [box] {}; 	\& \& \& \& \& \&  \node(0,0) (GPU 4) [box] {}; 	\&  \node(0,0) (GPU time 4) {}; \\
  \node(0,0) (CPU time 5) {}; 	\& \node(0,0) (CPU 5) [box] {}; 	\& \& \& \& \& \&  \node(0,0) (GPU 5) [box] {}; 	\&  \node(0,0) (GPU time 5) {}; \\
  \node(0,0) (CPU time 6) {}; 	\& \node(0,0) (CPU 6) [box] {};	\& \& \& \& \& \&  \node(0,0) (GPU 6) [box] {}; 	\&  \node(0,0) (GPU time 6) {}; \\
  \node(0,0) (CPU time 7) {}; 	\& \node(0,0) (CPU 7) [box] {};	\& \& \& \& \& \&  \node(0,0) (GPU 7) [box] {}; 	\&  \node(0,0) (GPU time 7) {}; \\
  \node(0,0) (CPU time 8) {}; 	\& \node(0,0) (CPU 8) [box] {};	\& \& \& \& \& \&  \node(0,0) (GPU 8) [box] {}; 	\&  \node(0,0) (GPU time 8) {}; \\
  \node(0,0) (CPU time 9) {}; 	\& \node(0,0) (CPU 9) [box] {};	\& \& \& \& \& \&  \node(0,0) (GPU 9) [box] {}; 	\&  \node(0,0) (GPU time 9) {}; \\
  \node(0,0) (CPU time 10) {};	\& \node(0,0) (CPU 10) [box] {};	\& \& \& \& \& \&  \node(0,0) (GPU 10) [box] {}; 	\&  \node(0,0) (GPU time 10) {}; \\
  \node(0,0) (CPU time 11) {}; 	\& \node(0,0) (CPU 11) [box] {}; 	\& \& \& \& \& \&  \node(0,0) (GPU 11) [box] {}; 	\&  \node(0,0) (GPU time 11) {}; \\
};

\fill 
	(CPU) node[draw,fill=white] (t1) {CPU}
	(GPU) node[draw,fill=white] (t2) {GPU};

\draw [dashed] (t1.south) -- (CPU 11.south);
\draw [dashed] (t2.south) -- (GPU 11.south);

\filldraw[fill=blue!30]
  (CPU 0.north west) rectangle (CPU 2.north east) 
  node (huffman) [midway,align=center] {Huffman 1\\(to GPU)};
\filldraw[fill=blue!30]
  (CPU 2.south west) rectangle (CPU 3.south east) 
  node (huffman) [midway,align=center] {Huffman 2\\(to GPU)};
\filldraw[fill=blue!30]
  (CPU 4.north west) rectangle (CPU 6.south east) 
  node (huffman) [midway,align=center] {Huffman 3\\(to GPU)};
\filldraw[fill=blue!20]
  (CPU 2.north west) rectangle (CPU 2.south east) 
  node (dispatch0) [text width=1.4cm, midway,align=center] {Dispatch};
\filldraw[fill=blue!20]
  (CPU 3.south west) rectangle (CPU 4.north east) 
  node (dispatch1) [text width=1.4cm, midway,align=center] {Dispatch};
\filldraw[fill=blue!20]
  (CPU 6.south west) rectangle (CPU 7.north east) 
  node (dispatch2) [text width=1.4cm, midway,align=center] {Dispatch};
\filldraw[fill=green!20]
  (GPU 2.north west) rectangle (GPU 4.north east) 
  node (opencl0) [midway,align=center] {GPU\\ (Huffman 1)};
\filldraw[fill=green!20]
  (GPU 4.north west) rectangle (GPU 6.north east) 
  node (opencl1) [midway,align=center] {GPU\\ (Huffman 2)};
\filldraw[fill=green!20]
  (GPU 6.south west) rectangle (GPU 8.south east) 
  node (opencl2) [midway,align=center] {GPU\\ (Huffman 3)};

\draw [-latex] (dispatch0.east) -- (dispatch0.east -| GPU 2.west) node [timelabel] (TextNode 0) {Write};
\draw [-latex] (dispatch1.east) -- (GPU 4.west) node [timelabel] (TextNode 1) {Write};
\draw [-latex] (dispatch2.east) -- (dispatch2.east -| GPU 8.west) node [timelabel] (TextNode 2) {Write};

\end{tikzpicture}
\label{fig:timelineGPU:b}}
\caption{JPEG decompression timelines: \protect\subref{fig:timelineGPU:a}
GPU execution of the parallel part after Huffman decoding and
\protect\subref{fig:timelineGPU:b}  pipelined execution of Huffman decoding and
GPU computations.  The CPU reads back the results at the end of each kernel
invocation.  The read arrows have been omitted for clarity.
\label{fig:timelineGPU}}
\end{center}
\end{figure}

We observed that Huffman decoding consumed around half of the overall execution
time with the SIMD-version of libjpeg-turbo.  Huffman decoding is sequential
and thus performed exclusively on the CPU. Subsequent decoding
steps, i.e., IDCT, upsampling and color conversions are highly
data-parallel and thus allocated to the GPU. In the following,
we refer to those steps as the {\em parallel part\/} of JPEG decoding.
 In the execution model explained so
far, GPU computations are delayed until decoded entropy data becomes available,
as shown in Figure~\ref{fig:timelineGPU:a}.  Because the GPU is un-utilized
during Huffman decoding, potential speedup is lost.  Using the fact that
entropy data is decoded in order, GPU computations can start after sufficient
image rows have been decoded.  Hence, Huffman decoding and GPU kernel execution
can be executed in a pipelined fashion, where the first pipeline stage, i.e.,
Huffman decoding, is executed on the CPU, and the second pipeline
stage, i.e., the parallel part, is executed on the GPU.
Figure~\ref{fig:timelineGPU:b} shows
the timeline of our pipelined execution mode. 

An image is sliced horizontally into several chunks.  As soon as the first
chunk is entropy decoded, the CPU transfers the data to the GPU.  All OpenCL
commands are executed asynchronously.  Hence, the CPU can resume Huffman
decoding immediately for the second chunk.  The execution time of the GPU
kernel may not match the Huffman decoding time, because Huffman decoding varies
greatly with the image details contained in a chunk.  If the GPU is faster than
the CPU, GPU computations are hidden behind Huffman-decoding on the CPU.  The
overall execution time then consists of the Huffman decoding time of the entire
image plus the kernel invocation for the last portion of the image. 

The most efficient chunk size is determined through static profiling on large
images.  Chunk sizes are varied from the full height down to an eight pixel
stripe.  The decoding speed tends to be faster as the number of chunks increases.
However, as chunks become too small, GPU utilization becomes low.
The best sizes from each image are selected.  The final
partition size is chosen as the largest size on the best list to prevent from
choosing a size that is too small wrt.~GPU utilization.

\section{Heterogeneous JPEG Decoding}
\label{sec:heterogjpeg}
Low-end GPUs may be incapable of out-performing high-end CPUs.  For such
CPU-GPU combinations, distributing the entire workload between GPU and CPU is
required.  We propose a performance model and partitioning scheme that
dynamically balances the workload on a CPU-GPU system.  We model execution time
based on an off-line profiling step.  This profiling is required only once for
a given CPU-GPU combination.  For profiling, we execute an instrumented version
of the JPEG decoder to determine the execution times of each decoding step for
a training set of images.  Multivariate polynomial regression analysis is
applied to derive closed forms that characterize the performance of a given
CPU-GPU combination.  We identified image entropy and the image dimensions as
the sole parameters for our performance model.  At run-time, the closed forms
are evaluated for a given image to estimate execution times and load-balance
the decoding workload between the CPU and the GPU.

\subsection{Performance Model}
\label{subsec:perfModel}

Our training set consists of twelve images from an online image
benchmark~\cite{RawzorImageBenchmark} and seven self-taken images.  Polynomial
regression poorly estimates performance for images with the dimensions outside
of the training set range.  Thus, the training-set baseline images are cropped
to create combinations of width and height up to
25~megapixels. The total number of images in the training set
is 4449. 

We categorize JPEG decoding stages into two phases: a sequential phase (Huffman
decoding) and a parallel phase (dequantization, IDCT, upsampling and color
conversion).  The sequential phase is executed exclusively on CPU while the
parallel phase can be executed on either the CPU or the GPU. Execution times
are collected for four decoding modes: sequential, SIMD, GPU and
pipelined GPU.  Execution time is measured using CPU timestamp counter
registers and the OpenCL event profiler.

\begin{figure}[htb]
    \begin{center}
        \includegraphics[clip=true]{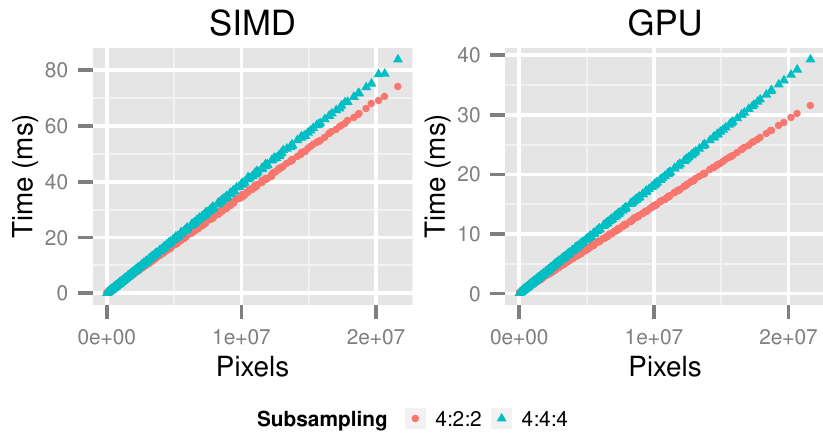}
    \caption{Execution time of SIMD and GPU of the parallel phase on GTX 560 scales linearly as image size increased. 
The other tested platforms showed a similar trend.
\label{fig:keneltime}}
    \end{center}
\end{figure}

\begin{figure}[htb]
    \begin{center}
        \includegraphics[clip=true]{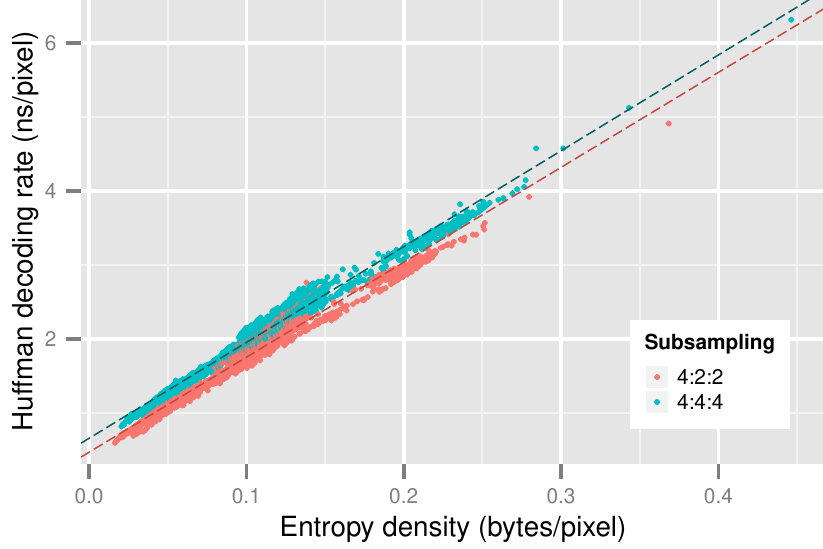}
    \end{center}
        \caption{Huffman decoding rate on GTX 560 with respect to the density of entropy in bytes per pixel along with best-fit lines.}
    \label{fig:huffmanExecution} 
\end{figure}

Figure~\ref{fig:keneltime} indicates that the parallel phase scales linearly
with respect to image size.  Thus, we perform polynomial fits on the parallel
phase, $\PCPU$ and $\PGPU$, as a function of image width and height.   
Huffman decoding does not show a linear relationship with image dimension.
We have observed that it varies on the complexity of the chrominance and
luminance of an image, which reflects on entropy size.
Figure~\ref{fig:huffmanExecution} suggests a linear relationship between
Huffman decoding time per pixel and entropy density.  Because the encoded bitstream
occupies the largest portion of a JPEG file, the density of entropy coded data
can be approximated from image file size and image dimensions as
\begin{equation}
	d = \frac{\mathit{ImageFileSize}}{w*h},
\end{equation}
where $w$ is the image width, $h$ is the image height and $d$ is the image's
entropy density per pixel.  We model the Huffman decoding rate,
$T_{\mathit{HuffmanPerPixel}}$, using polynomial regression as a function of
entropy density.  The Huffman decoding time of the entire image, $\THuffman$, is
approximated as follows.
\begin{equation}
\label{equation:huffman}
\THuffman(w,h,d) = T_{\mathit{HuffmanPerPixel}}(d) * w * h
\end{equation}
This equation assumes that entropy data is evenly distributed across an image, which
we found to be a workable approximation.

The variables to our performance model are image width, height and entropy data
size.  We model each phase using polynomial regression up to a degree of seven.
The best fit model is selected by comparing Akaike information
criteria~\cite{Akaike19813}.  Modelling with higher degrees is computationally
possible.  However, we have observed that higher degrees do not imply a more
precise model, and performance may suffer from the higher prediction time
required to evaluate polynomials of higher degrees.

Evaluating polynomials of high degrees at run-time showed a noticeable negative
impact on the performance of the JPEG decoder.  We rearranged all polynomials
in Horner form~\cite{horner1819} to reduce the number of multiplications
required for polynomial evaluations. With this optimization the prediction
overhead became negligible compared to the overall execution time for decoding.

The overall execution time on a CPU can be expressed as a summation of the sequential phase and the parallel phase:
\begin{equation}
 \TTotal = \THuffman(w,h,d) + \PCPU(w,h),
\end{equation}

When we profile execution times on the GPU, OpenCL work-group sizes are
alternated from 4 MCUs to 32 MCUs to find the best work-group size for a
specific platform.  Similar to the CPU model, the total execution time for the
GPU mode is expressed a summation of Huffman decoding time and GPU execution
time.
\begin{equation}
 \TTotal = \THuffman(w,h,d) + \PGPU(w,h)
\end{equation}
Data transfers between CPU and GPU device generate significant
overhead~\cite{GPUtransfer2011}.  The time collected for the GPU includes
data transfer overhead and
the GPU kernel computation,
\begin{equation}
\label{gpubeakdown}
 \PGPU(w,h) = O_{w}(w,h) + T_{kernel}(w,h) + O_{r}(w,h),
\end{equation}
where $O_{w}$ and $O_{r}$ are data transfer costs from the CPU to the
GPU and vice versa.
The input and output buffers are pinned, to achieve faster transfers~\cite{OpenCLProgrammingGuide4.2}.

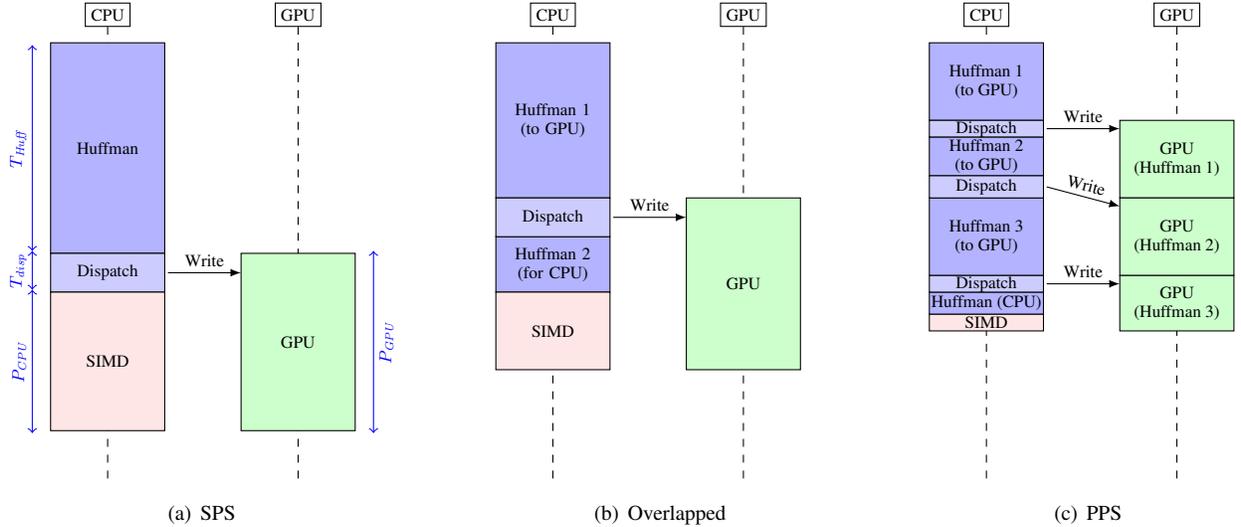
\begin{figure*}[htb]
\begin{center}
\subfigure[SPS]{
\begin{tikzpicture}[
  every node/.style={font=\normalsize,
  minimum height=0.30cm,minimum width=0.3cm},
  box/.style={minimum height=0.30cm,minimum width=2.1cm},
  timelabel/.style={font=\normalsize,midway, above, sloped},
  scale=0.75,every node/.style={scale=0.72},
  ]

\node [matrix, ampersand replacement=\&, very thin,column sep=0.17cm,row sep=0.30cm] (matrix) at (0,0) {
  \node(0,0) (duration) {}; \& \node(0,0) (CPU) {}; 	\& \& \& \& \& \&
  \node(0,0) (GPU) {}; \&  \node(0,0) (GPU time) {}; \\  
  \node(0,0) (CPU time 0) {}; \& \node(0,0) (CPU 0) [box] {};	 \& \& \& \& \& \&
  \node(0,0) (GPU 0) [box] {}; \&  \node(0,0) (GPU time 0) {}; \\
  \node(0,0) (CPU time 1) {}; \&  \node(0,0) (CPU 1) [box] {};	\& \& \& \& \& \&
  \node(0,0) (GPU 1) [box] {}; \&  \node(0,0) (GPU time 1) {}; \\
  \node(0,0) (CPU time 2) {}; \&  \node(0,0) (CPU 2) [box] {};  \& \& \& \& \& \&
  \node(0,0) (GPU 2) [box] {}; \&  \node(0,0) (GPU time 2) {};  \\
  \node(0,0) (CPU time 3) {}; \&  \node(0,0) (CPU 3) [box] {};	\& \& \& \& \& \&
  \node(0,0) (GPU 3) [box] {}; \&  \node(0,0) (GPU time 3) {};  \\
  \node(0,0) (CPU time 4) {}; \&  \node(0,0) (CPU 4) [box] {}; 	\& \& \& \& \& \&
  \node(0,0) (GPU 4) [box] {}; \&  \node(0,0) (GPU time 4) {};  \\
  \node(0,0) (CPU time 5) {}; \&  \node(0,0) (CPU 5) [box] {}; 	\& \& \& \& \& \&
  \node(0,0) (GPU 5) [box] {}; \&  \node(0,0) (GPU time 5) {};  \\
  \node(0,0) (CPU time 6) {}; \&  \node(0,0) (CPU 6) [box] {};	\& \& \& \& \& \&
  \node(0,0) (GPU 6) [box] {}; \&   \node(0,0) (GPU time 6) {};  \\
  \node(0,0) (CPU time 7) {}; \& \node(0,0) (CPU 7) [box] {};	 \& \& \& \& \& \&
  \node(0,0) (GPU 7) [box] {}; \&  \node(0,0) (GPU time 7) {}; \\
  \node(0,0) (CPU time 8) {}; \&  \node(0,0) (CPU 8) [box] {};	\& \& \& \& \& \&
  \node(0,0) (GPU 8) [box] {}; \&  \node(0,0) (GPU time 8) {}; \\
  \node(0,0) (CPU time 9) {}; \&  \node(0,0) (CPU 9) [box] {};  \& \& \& \& \& \&
  \node(0,0) (GPU 9) [box] {}; \&  \node(0,0) (GPU time 9) {};  \\
  \node(0,0) (CPU time 10) {}; \&  \node(0,0) (CPU 10) [box] {};	\& \& \& \& \& \&
  \node(0,0) (GPU 10) [box] {}; \&  \node(0,0) (GPU time 10) {};  \\
  \node(0,0) (CPU time 11) {}; \&  \node(0,0) (CPU 11) [box] {}; 	\& \& \& \& \& \&
  \node(0,0) (GPU 11) [box] {}; \&  \node(0,0) (GPU time 11) {};  \\
};

\fill 
	(CPU) node[draw,fill=white] (t1) {CPU}
	(GPU) node[draw,fill=white] (t2) {GPU};

\draw [dashed] (t1.south) -- (CPU 11.south);
\draw [dashed] (t2.south) -- (GPU 11.south);

\filldraw[fill=blue!30]
  (CPU 0.north west) rectangle (CPU 5.south east) node (huffman) [midway,align=center] {Huffman};
\filldraw[fill=blue!20]
  (CPU 5.south west) rectangle (CPU 6.south east) node (dispatch) [text width=2.0cm, midway,align=center] {Dispatch};
\filldraw[fill=green!20]
  (GPU 5.south west) rectangle (GPU 10.north east) node (opencl) [midway,align=center] {GPU};
\filldraw[fill=red!10]
  (CPU 6.south west) rectangle (CPU 10.north east) node (simd)
[midway,align=center] {SIMD};

\draw[<->,blue]  (CPU time 5.south) -- (CPU time 0.north) 
     node [timelabel] (TextNode 0) {$\THuffman$};
\draw[<->,blue] (CPU time 6.south) -- (CPU time 5.south)  
     node [timelabel] (TextNode 1) {$\TDispatch$};
\draw[<->,blue]  (GPU time 10.north) -- (GPU time 5.south) 
     node [timelabel,below] (TextNode 2) {$\PGPU$};
\draw[<->,blue] (CPU time 10.north) -- (CPU time 6.south)  
     node [timelabel] (TextNode 1) {$\PCPU$};     

\draw [-latex] (dispatch.east) -- (dispatch.east -| GPU 4.west) node [timelabel] (TextNode) {Write};
\end{tikzpicture}
\label{fig:timelineHeterog:a}
}
\qquad\quad
\subfigure[Overlapped]{
\begin{tikzpicture}[
  every node/.style={font=\normalsize,
  minimum height=0.30cm,minimum width=0.3cm},
  box/.style={minimum height=0.30cm,minimum width=2.1cm},
  timelabel/.style={font=\normalsize,midway, above, sloped},
  scale=0.75,every node/.style={scale=0.72},
  ]

\node [matrix, ampersand replacement=\&, very thin,column sep=0.17cm,row sep=0.30cm] (matrix) at (0,0) {
  \node(0,0) (CPU) {}; 	\& \& \& \& \& \&
  \node(0,0) (GPU) {}; \&  \node(0,0) (GPU time) {}; \\  
  \node(0,0) (CPU 0) [box] {};	\& \& \& \& \& \&  \node(0,0) (GPU 0) [box] {};\\
  \node(0,0) (CPU 1) [box] {};	\& \& \& \& \& \&  \node(0,0) (GPU 1) [box] {};\\
  \node(0,0) (CPU 2) [box] {};   \& \& \& \& \& \&  \node(0,0) (GPU 2) [box] {};\\
  \node(0,0) (CPU 3) [box] {};	\& \& \& \& \& \&  \node(0,0) (GPU 3) [box] {};\\
  \node(0,0) (CPU 4) [box] {}; 	\& \& \& \& \& \&  \node(0,0) (GPU 4) [box] {};\\
  \node(0,0) (CPU 5) [box] {}; 	\& \& \& \& \& \&  \node(0,0) (GPU 5) [box] {};\\
  \node(0,0) (CPU 6) [box] {};	\& \& \& \& \& \&  \node(0,0) (GPU 6) [box] {};\\
  \node(0,0) (CPU 7) [box] {};	\& \& \& \& \& \&  \node(0,0) (GPU 7) [box] {};\\
  \node(0,0) (CPU 8) [box] {};	\& \& \& \& \& \&  \node(0,0) (GPU 8) [box] {};\\
  \node(0,0) (CPU 9) [box] {};  \& \& \& \& \& \&  \node(0,0) (GPU 9) [box] {};\\
  \node(0,0) (CPU 10) [box] {};	\& \& \& \& \& \&  \node(0,0) (GPU 10) [box] {};\\
  \node(0,0) (CPU 11) [box] {}; 	\& \& \& \& \& \&  \node(0,0) (GPU 11) [box] {};\\
};

\fill 
	(CPU) node[draw,fill=white] (t1) {CPU}
	(GPU) node[draw,fill=white] (t2) {GPU};

\draw [dashed] (t1.south) -- (CPU 11.south);
\draw [dashed] (t2.south) -- (GPU 11.south);

\filldraw[fill=blue!30]
  (CPU 0.north west) rectangle (CPU 4.north east) node (huffman) [text width=2.0cm, midway,align=center] {Huffman 1\\ (to GPU)};
  \filldraw[fill=blue!30]
  (CPU 5.north west) rectangle (CPU 6.south east) node (huffman) [text width=2.0cm, midway,align=center] {Huffman 2\\ (for CPU)};
\filldraw[fill=blue!20]
  (CPU 4.north west) rectangle (CPU 5.north east) node (dispatch) [text width=2.0cm, midway,align=center] {Dispatch};
\filldraw[fill=green!20]
  (GPU 4.north west) rectangle (GPU 8.south east) node (opencl) [text width=2.0cm, midway,align=center] {GPU};
\filldraw[fill=red!10]
  (CPU 6.south west) rectangle (CPU 8.south east) node (simd)
[midway,align=center] {SIMD};

\draw [-latex] (dispatch.east) -- (dispatch.east -| GPU 4.west) node [timelabel] (TextNode) {Write};
\end{tikzpicture}
\label{fig:timelineHeterog:b}}
\qquad\quad
\subfigure[PPS]{
\begin{tikzpicture}[
  every node/.style={font=\normalsize,
  minimum height=0.30cm,minimum width=0.3cm},
  box/.style={minimum height=0.30cm,minimum width=2.1cm},
  timelabel/.style={font=\normalsize,midway, above, sloped},
  scale=0.75,every node/.style={scale=0.72},
  ]

\node [matrix,ampersand replacement=\&, very thin,column sep=0.17cm,row sep=0.30cm] (matrix) at (0,0) {
  \node(0,0) (CPU) {}; 	\& \& \& \& \& \&  \node(0,0) (GPU) {}; \&\\  
  \node(0,0) (CPU 0) [box] {};	\& \& \& \& \& \&  \node(0,0) (GPU 0) [box] {};\\
  \node(0,0) (CPU 1) [box] {};	\& \& \& \& \& \&  \node(0,0) (GPU 1) [box] {};\\
  \node(0,0) (CPU 2) [box] {};  	\& \& \& \& \& \&  \node(0,0) (GPU 2) [box] {};\\
  \node(0,0) (CPU 3) [box] {};	\& \& \& \& \& \&  \node(0,0) (GPU 3) [box] {};\\
  \node(0,0) (CPU 4) [box] {}; 	\& \& \& \& \& \&  \node(0,0) (GPU 4) [box] {};\\
  \node(0,0) (CPU 5) [box] {}; 	\& \& \& \& \& \&  \node(0,0) (GPU 5) [box] {};\\
  \node(0,0) (CPU 6) [box] {};	\& \& \& \& \& \&  \node(0,0) (GPU 6) [box] {};\\
  \node(0,0) (CPU 7) [box] {};	\& \& \& \& \& \&  \node(0,0) (GPU 7) [box] {};\\
  \node(0,0) (CPU 8) [box] {};	\& \& \& \& \& \&  \node(0,0) (GPU 8) [box] {};\\
  \node(0,0) (CPU 9) [box] {};  	\& \& \& \& \& \&  \node(0,0) (GPU 9) [box] {};\\
  \node(0,0) (CPU 10) [box] {};	\& \& \& \& \& \&  \node(0,0) (GPU 10) [box] {};\\
  \node(0,0) (CPU 11) [box] {}; 	\& \& \& \& \& \&  \node(0,0) (GPU 11) [box] {};\\
};

\fill 
	(CPU) node[draw,fill=white] (t1) {CPU}
	(GPU) node[draw,fill=white] (t2) {GPU};

\draw [dashed] (t1.south) -- (CPU 11.south);
\draw [dashed] (t2.south) -- (GPU 11.south);

\filldraw[fill=blue!30]
  (CPU 0.north west) rectangle (CPU 2.north east) 
  node (huffman) [midway,align=center] {Huffman 1\\ (to GPU)};
\filldraw[fill=blue!30]
  (CPU 2.south west) rectangle (CPU 3.south east) 
  node (huffman) [midway,align=center] {Huffman 2\\ (to GPU)};
\filldraw[fill=blue!30]
  (CPU 4.north west) rectangle (CPU 6.north east) 
  node (huffman) [midway,align=center] {Huffman 3\\ (to GPU)};
\filldraw[fill=blue!30]
  (CPU 6.south west) rectangle (CPU 7.north east) 
  node (huffman) [midway,align=center] {Huffman (CPU)};
\filldraw[fill=blue!20]
  (CPU 2.north west) rectangle (CPU 2.south east) 
  node (dispatch0) [text width=2.0cm,midway,align=center] {Dispatch};
\filldraw[fill=blue!20]
  (CPU 3.south west) rectangle (CPU 4.north east) 
  node (dispatch1) [text width=2.0cm,midway,align=center] {Dispatch};
\filldraw[fill=blue!20]
  (CPU 6.north west) rectangle (CPU 6.south east) 
  node (dispatch2) [text width=2.0cm,midway,align=center] {Dispatch};
\filldraw[fill=green!20]
  (GPU 2.north west) rectangle (GPU 4.north east) 
  node (opencl0) [midway,align=center] {GPU\\ (Huffman 1)};
\filldraw[fill=green!20]
  (GPU 4.north west) rectangle (GPU 6.north east) 
  node (opencl1) [midway,align=center] {GPU\\ (Huffman 2)};
\filldraw[fill=green!20]
  (GPU 6.north west) rectangle (GPU 7.south east) 
  node (opencl2) [midway,align=center] {GPU\\ (Huffman 3)};
\filldraw[fill=red!10]
  (CPU 7.north west) rectangle (CPU 7.south east) node (simd)
[midway,align=center] {SIMD};

\draw [-latex] (dispatch0.east) -- (dispatch0.east -| GPU 2.west) node [timelabel] (TextNode 0) {Write};
\draw [-latex] (dispatch1.east) -- (GPU 4.west) node [timelabel] (TextNode 1) {Write};
\draw [-latex] (dispatch2.east) -- (dispatch2.east -| GPU 6.west) node [timelabel] (TextNode 2) {Write};
\end{tikzpicture}
\label{fig:timelineHeterog:c}}
\end{center}
\vspace{-4mm}
\caption{Heterogeneous JPEG decoding timelines of three execution models:
\protect\subref{fig:timelineHeterog:a} SPS,
\protect\subref{fig:timelineHeterog:b} overlapped Huffman decoding and GPU
execution, and \protect\subref{fig:timelineHeterog:c} PPS.
Partitioning schemes are depicted proportionally; PPS achieves the highest
overlap of GPU kernel execution with the sequential Huffman decoding on the
CPU. Note that the last
GPU invocation is shorter because the work is shared with the CPU (SIMD).
\label{fig:timelineHeterog}}
\end{figure*}

\subsection{Partitioning Schemes}
For reasons introduced in Section~\ref{sec:intro}, Huffman decoding constitutes
the non-parallelizable part of JPEG decoding, which is thus entirely executed on
the CPU. Subsequent decoding steps, i.e., IDCT, upsampling and color conversion
constitute the parallelizable part for which we utilize both the CPU and the GPU. 
For the parallelizable part, our partitioning scheme splits images horizontally such that the
initial $x$~rows of the image are assigned to the GPU, and the remaining
$h-x$~rows are assigned to the CPU. The value for variable~$x$ is chosen
such that the overall execution times for the CPU and GPU are equal, i.e., the load
is equally balanced. Variable~$x$ is rounded to the nearest value evenly divisible
by the number of rows in an MCU. This requirement is due to libjpeg-turbo's
convention to decode images in units of MCUs.
The input parameters to our partitioning schemes are the image dimensions and the
image entropy, approximated by bytes/pixel derived from the image data size and
the image dimensions.

\subsubsection{Simple Partitioning Scheme (SPS):}
The simplest approach is to parallelize the computations after Huffman
decoding.  CPU and GPU perform the parallel phase concurrently.
Figure~\ref{fig:timelineHeterog:a} illustrates the SPS partitioning scheme.
The CPU first performs entropy decoding of the entire image, then partitions the resulting
image data in two parts.  The first part is processed by the GPU and the second part by
the CPU.  Data transfer commands between CPU and GPU and kernel launching
commands are asynchronous
calls.  Hence, the CPU is allowed to continue execution after dispatching
commands to the GPU.
The overall execution time can be modelled as the maximum time of the two architectures.
\begin{equation}
\TTotal = max(\TCPU,\TGPU)
\end{equation}
With the SPS model, the CPU execution time, $\TCPU$, and the GPU execution time, $\TGPU$,
are expressed as
{\small
\begin{subequations}
\begin{align}
\TCPU(w,h) &= \THuffman(w,h,d) +,\TDispatch(w,h-x) + \PCPU(w,x), \text{ and}\\
\TGPU(w,h) &= \THuffman(w,h,d) + \PGPU(w,h-x),
\end{align}
\end{subequations}
}
where $x$ is the number of rows assigned to the CPU, and $\TDispatch$ is the
amount
of time the CPU spends on the OpenCL kernel invocation.  $\PGPU$ includes kernel 
execution time and data transfer overhead. The workload is
considered well-balanced when the parallel parts on both architectures achieve
the same execution time.
{\small
\begin{equation}
f(x) = \TDispatch(w,h-x) + \PCPU(w,x) - \PGPU(w,h-x)\label{equation:simpleHeterog}
\end{equation}
}
The only unknown variable is the number of rows assigned to the CPU, i.e.,~$x$. 
When $f(x)$ become zero, the execution time is balanced.
This problem is equivalent to the root solving problem. 
At run-time, the root can be estimated using Newton's method,
\begin{equation}
x_{n+1} = x_n - \frac{f(x_n)}{f'(x_n)},
\end{equation}
where $x_n$ is the initial partitioning height to the CPU, $x_{n+1}$ is the new
height approximation, $f(x_n)$ is given as
Equation~\eqref{equation:simpleHeterog}, and $f'(x_n)$ is the first derivation.
Newton's method is performed recursively until no better partition can be
found.  The GPU computes the parallel phase on the sub-image of size $w$ by
$h-x$, while the CPU computes the remaining $x$ image rows.

\subsubsection{Pipelined Partitioning Scheme (PPS):}
The GPU is underutilized during the Huffman decoding stage in SPS.
We have demonstrated in Section~\ref{subsec:pipelinedGPU} that sequential
Huffman decoding can be parallelized with GPU kernel execution.  Entropy data
for the CPU can be decoded simultaneously with GPU computations as illustrated in
Figure~\ref{fig:timelineHeterog:b}.  The concurrent execution happens after the
Huffman decoding part for the GPU.  The total execution time on each architecture can be modelled as follows.
{\small
\begin{subequations}
\begin{align}
\TCPU &= \THuffman(w,h,d) + \PCPU(w,x) + \TDispatch(w,h-x)\\
\TGPU &= \THuffman(w,h-x,d) + \PGPU(w,h-x)
\end{align}
\end{subequations}
}
This partitioning scheme balances GPU execution with the sum of OpenCL
dispatching time, entropy decoding time for the CPU part and the computation
time of the parallel part on the CPU.
{\small
\begin{equation}
\label{equation:overlap}
\begin{split}
f(x) = &\TDispatch(w,h-x) + \THuffman(w,h,d) + \PCPU(w,x) \\
&- \PGPU(w,h-x)
\end{split}
\end{equation}
}
Similar to SPS, we use Newton's method to approximate variable~$x$ at run-time.

Further parallelism can be achieved by 
pipelining the GPU kernel executions with this partitioning scheme.
Figure~\ref{fig:timelineHeterog:c} shows an
execution timeline of PPS.  The parallel part for the first image chunk
(Huffman 1)
is started on the GPU immediately after it has been transferred from
the CPU.
{\small
\begin{subequations}
\begin{align}
\TCPU &= \THuffman(w,h,d) + \PCPU(w,x) + \TDispatch(w,h-x)\\
\TGPU &= \THuffman(w,c,d) + \PGPU(w,h-x)
\end{align}
\end{subequations}
}
In the above equation, 
$c$ is the number of image rows per chunk decoded on the GPU. 
The chunk size has been determined through profiling as explained in Section~\ref{subsec:pipelinedGPU}. 
The number of rows per chunk can be calculated by chunk size divided by image width. 
Therefore, the partitioning equation becomes 
{\small
\begin{equation}
\begin{split}
f(x) = &\THuffman(w,h-c,d) + \PCPU(w,x) + \TDispatch(w,h-x)\\
&- \PGPU(w,h-x),
\end{split}
\end{equation}
}
where $h-c$ denotes the remaining rows in an image after the first chunk.

Our Huffman decoding time estimation model assumes a uniform entropy distribution
across an image.  However, the density of entropy data is unlikely to be evenly
distributed in practice.  Using the Huffman decoding time model to estimate the time
for a certain chunk is often imprecise because the average
density and the actual density of the chunk are mismatched.  We compensate the
error by re-partitioning.  Throughout the computation, we keep records of the
actual Huffman decoding times.  Before entropy decoding of the last GPU chunk,
workload distribution is re-calculated. 
At this point, one GPU chunk and the CPU partition remain unprocessed. 
Thus, a modification of Equation~\eqref{equation:overlap} can be used,
{\small
\begin{equation}
\label{equation:re-calculation}
\begin{split}
f(x) = &\TDispatch(w,h'-x') + \THuffman(w,h',d') + \PCPU(w,x') \\
 &- \PGPU(w,h-x) - P_{prevGPU},
 \end{split}
\end{equation}
}
where $h'$ is the unprocessed height, $x'$ is the new height allocated to the GPU,
and $d'$ is the new density rate.  The previous kernel execution may not
complete by the time of re-partitioning.  $T_{prevGPU}$ is an estimated
remaining time from the previous GPU kernel invocation that potentially
influences the new partitioning scheme.  The remaining height, $h'$, is known at
run-time.  The only unknown variable to be solved at run-time, is $x'$.

We estimate the total Huffman decoding time using
Equation~\eqref{equation:huffman}, and the actual Huffman decoding time of
previous chunks are known at run-time.  Figure~\ref{fig:huffmanExecution}
implies a linear proportional relationship between an image dimension and
Huffman decoding time.  The new density is calculated using the ratio of the
remaining Huffman time and the image height.
\begin{equation}
d' = \frac{\mathit{HuffmanDecodingTimeRatio}}{\mathit{ImageHeightRatio}} * d,
\end{equation}
Above, $HuffmanDecodingTimeRatio$ is the ratio of the remaining decoding
time to the estimated total decoding time, and $ImageHeightRatio$ is the
ratio of unprocessed height to the total image height.  When the ratio of the
remaining Huffman time is greater than the height ratio, the remaining part of
an image consists of more detail.  It indicates that the entropy data rate
becomes denser and more workload should be allocated to the GPU.  Otherwise, the
entropy data rate becomes less dense, and more workload should be allocated to
the CPU. 

Even though Figure~\ref{fig:timelineHeterog:c} shows a small gain compared 
to Figure~\ref{fig:timelineGPU:b}, this approach actually yields 
large improvements on a hardware configuration where the CPU is more powerful
than the GPU.

\section{Experimental Results}
\label{sec:experiment}
We conducted an extensive experimental evaluation on six versions of the JPEG
decoder, namely the sequential version, SIMD, GPU, pipelined GPU, SPS and PPS,
on three representative platforms specified in Table~\ref{tab:spec}.  All file
I/O instructions were disabled to minimize time variations that do not reflect
the actual performance of the algorithm.  To demonstrate the effectiveness of
our implementation, we used two chroma subsamplings,
i.e.,
4:2:2 and 4:4:4.  The other subsamplings are decoded in a similar manner as
4:2:2 images. 
For the performance evaluation we used a new set of images that does not share
any images with the training set.  Our image test-set consists of fourteen
images from {CorpusNielsFrohling} and three self-taken images.  These images
are cropped to various sizes summing up to the total of 3597 images for each
subsampling. 

\begin{table}[h!btp]
\centering
\small
\begin{tabular}{|l|c|c|c|}
\hline
\textbf{Machine name} & \textbf{GT 430} & \textbf{GTX 560} & \textbf{GTX 680}\\ \hline \hline
CPU model & Intel i7-2600k & Intel i7-2600k & Intel i7-3770k \\ \hline
CPU frequency & 3.4 GHz & 3.4 GHz & 3.5 GHz \\ \hline
No. of CPU cores & 4 & 4 & 4\\ \hline \hline
GPU model & \begin{tabular}[x]{@{}c@{}}NVIDIA\\GT 430\end{tabular} 
& \begin{tabular}[x]{@{}c@{}}NVIDIA\\GTX 560Ti\end{tabular} 
& \begin{tabular}[x]{@{}c@{}}NVIDIA\\GTX 680\end{tabular} \\ \hline
GPU core frequency & 700 MHz & 822MHz & 1006MHz \\ \hline
No. of GPU cores & 96 & 384 & 1536 \\ \hline
GPU memory size & 1024 MB & 1024 MB & 2048 MB \\ \hline
Compute Capability & 2.1 & 2.1 & 3.0 \\ \hline \hline
Ubuntu version  & 11.04 & 12.04  & 11.04 \\ \hline
Linux Kernel & 2.6.38 & 3.5.0 & 2.6.38 \\ \hline
GCC version& 4.5.2 & 4.6.3 & 4.5.2 \\ \hline
\end{tabular}
\caption{Hardware Specifications.}
\label{tab:spec}
\end{table}
%
\subsection{OpenCL Kernel Execution}
\begin{figure}[htb]
    \begin{center}
        \includegraphics[clip=true]{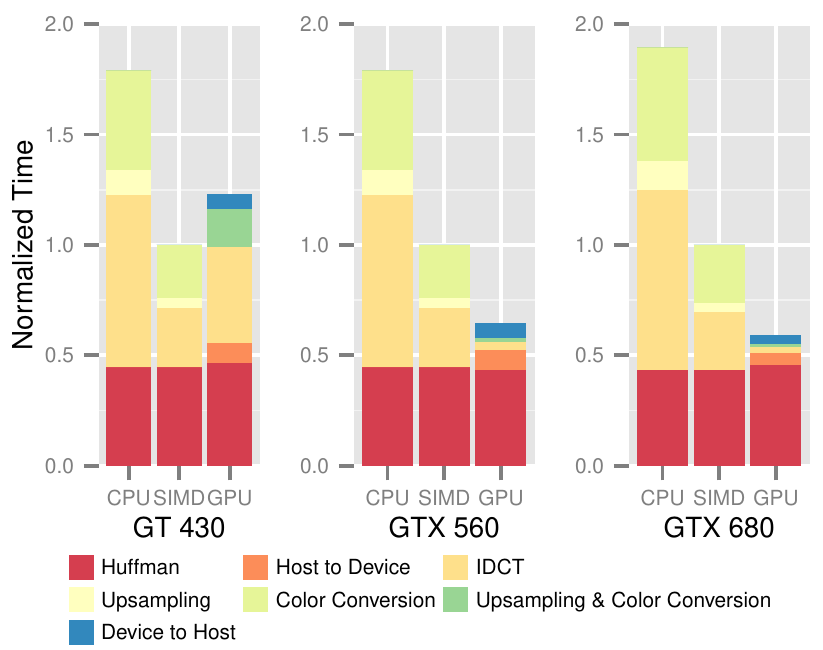}
    \caption{Decoding time normalized with respect to JPEG decompression in
SIMD mode.  The decoded image's dimension is 2048x2048 with 4:2:2 subsampling.
Shown are the execution time break-downs of libjpeg-turbo's sequential JPEG decoder
 on the CPU, SIMD
execution on a CPU with libjpeg-turbo, and our GPU execution.
    \label{fig:GPUBar}}
    \end{center}
\end{figure}

Figure~\ref{fig:GPUBar} depicts a break-down of execution times
for the sequential CPU,
SIMD and GPU modes.  The y-axis is normalized with respect to
SIMD execution times.  Our GPU computations on all tested architectures surpass
the sequential CPU execution.  For this specific image size, the GPU
computation was able to reduce 35.5\% and 40.8\% of the overall execution time
on GTX 560 and GTX 680 respectively.  The kernel execution from IDCT to 
color conversion was 10x faster than the SIMD execution on the GTX 560 and 13.7x
faster on the GTX 680.  However, taking data transfer overhead into account, the
performance improvements were  reduced to 2.6x and 4.3x.

It follows from Fig.~\ref{fig:GPUBar}
that performance improvements are not guaranteed by migrating computations exclusively
to a GPU.  GT~430, consisting of 96 cores, is the weakest GPU among the three
representative machines.  The experimental result on the 2048x2048 image showed
a 23\% slow-down compared to SIMD execution on an Intel i7.  The kernel
execution shows a 27\% slower data transfer between CPU and GPU.


\subsection{Heterogeneous JPEG Decoding Performance}
\begin{figure}[htb]
    \begin{center}
        \includegraphics[clip=true]{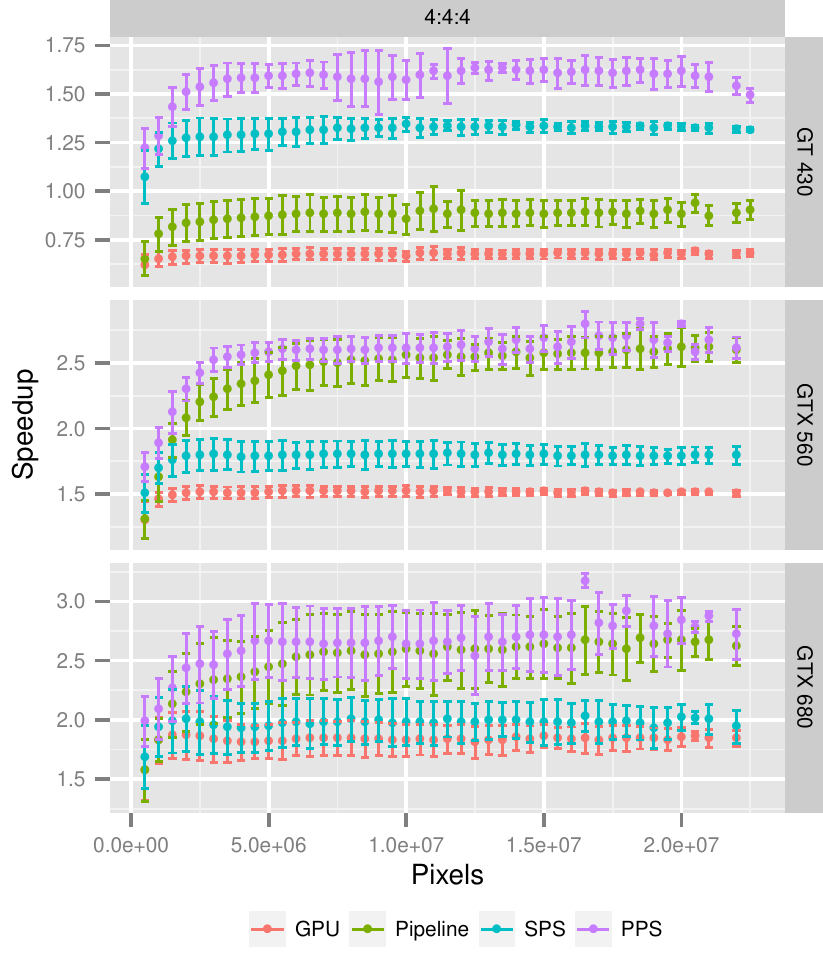}
    \caption{Average speedups over libjpeg-turbo's SIMD execution with respect
to image size in pixels on the three representative machines. The error bar
represents standard deviation.  \label{fig:Speedup}}
    \end{center}
\end{figure}

We evaluated our heterogeneous JPEG decompression models with respect to the
SIMD-version of libjpeg-turbo.  Figure~\ref{fig:Speedup} shows the average speedups
with standard deviation bars as image size increased.  Due to space
limitations, we only provide the results for 4:4:4 subsampling.  A similar
trend was observed for 4:2:2 subsampling.  Table~\ref{tab:speedup422} and
Table~\ref{tab:speedup444} summarize the performances of 4:2:2 and 4:4:4
subsampling respectively. 

\begin{table}[h!tbp]
\centering
\small
\begin{tabular}{|l|c|c|c|}
\hline
\textbf{Mode} & \textbf{GT 430} & \textbf{GTX 560} & \textbf{GTX 680}\\ \hline
GPU & $0.72 \pm 5.35\%$ & $1.59 \pm 7.50\%$ & $1.94 \pm 12.94\%$ \\ \hline
Pipeline & $0.92 \pm 13.67\%$ & $2.19 \pm 20.21\%$ & $2.33 \pm 20.75\%$ \\ \hline
SPS & $1.31 \pm 9.54\%$ & $1.81 \pm 10.13\%$ & $2.04 \pm 15.15\%$ \\ \hline
PPS & $1.54 \pm 10.93\%$ & $2.34 \pm 15.19\%$ & $2.52 \pm 17.08\%$ \\ \hline
\end{tabular}
\caption{Average speedup and coefficient of variation over SIMD execution when decoding 4:2:2 subsampled images.}
\label{tab:speedup422}
\end{table}

\begin{table}[h!tbp]
\centering
\small
\begin{tabular}{|l|c|c|c|}
\hline
\textbf{Mode} & \textbf{GT 430} & \textbf{GTX 560} & \textbf{GTX 680}\\ \hline
GPU & $0.66 \pm 5.82\%$ & $1.49 \pm 5.87\%$ & $1.81 \pm 10.84\%$ \\ \hline
Pipeline & $0.83 \pm 13.48\%$ & $2.14 \pm 19.97\%$ & $2.26 \pm 19.48\%$ \\ \hline
SSP & $1.27 \pm 8.62\%$ & $1.76 \pm 8.12\%$ & $1.94 \pm 12.55\%$ \\ \hline
PPS & $1.50 \pm 10.46\%$ & $2.34 \pm 14.33\%$ & $2.45 \pm 15.02\%$ \\ \hline
\end{tabular}
\caption{Average speedup and coefficient of variation over SIMD execution when decoding 4:4:4 subsampled images.}
\label{tab:speedup444}
\end{table}

PPS achieves the highest performance on all machines.  It attains average speedups
of 1.5x, 2.3x and 2.5x over SIMD mode and 3.1x, 4.8x and 5.2x over sequential
execution on GTX 430, GTX 560 and GTX 680 respectively.  The highest-recorded
speedups were 4.2x faster than SIMD and 8.5x faster than sequential execution
on GTX 680.  PPS does not show a signification improvement over pipelined GPU
execution on GTX 560 and GTX 680 because most GPU kernel executions were
sufficiently fast to hide within the Huffman decompression time.  Therefore,
only a small amount of workload was allocated to the CPU, and a small improvement
was achieved.

On GT 430, the GPU mode and the pipelined GPU execution mode failed to surpass
SIMD.  As a result, both of our partitioning schemes distributed the larger
partition to the CPU.  Despite the slow GPU, the cooperative CPU-GPU
executions achieved speedups over libjpeg-turbo's SIMD mode. 
 
The pipelined execution is always faster than a single large GPU kernel
invocation because entropy is decoded simultaneously with a GPU computation to
reduce the hardware idle time.  When the decoded image has a size smaller than
the pre-determined chunk size, the image is executed as one GPU kernel invocation.
Therefore, no improvement is shown over the normal GPU mode. 

It should be noted that the GTX 680 has larger coefficients of variation than the
other machines.  This fluctuation reflected the contribution of Huffman
decoding time to the speedup calculation.  An image with larger entropy data takes
longer time to decode.  As a result, the overall speedup becomes smaller than an
image with sparser entropy.  The faster GPU is more sensitive to the change.
Therefore, GTX 680 suffered the highest impact from a small change in Huffman
decoding time compared to the other tested machines.
%



\begin{figure}[htb]
    \begin{center}
        \includegraphics[clip=true]{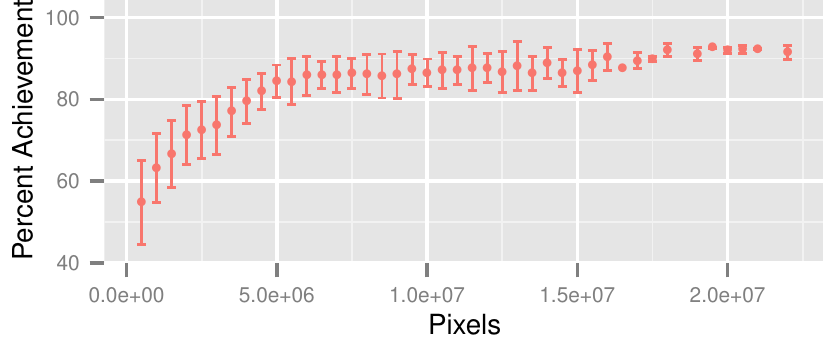}
    \caption{Speedup Comparison of PPS execution to the maximum achievable speedup on GTX 680.
The points represent the mean percent achieved along with standard deviation bars.
\label{fig:pcnt_achvd}}
    \end{center}
\end{figure}

According to Amdahl's law, the maximum attainable speedup is restricted by
the sequential portion of the program.  Equation~\eqref{equation:amdahl} 
states the theoretical speedup assuming an infinite number of processors. 
\begin{equation}
\label{equation:amdahl}
	\mathit{Speedup} = \frac{\TTotal}{\TTotal*(1-P)}
\end{equation}
$P$ is the fraction of the parallelizable portion of the program, and $1-P$ is
the serial portion, which in this case, is entropy (Huffman) decoding. 
Thus, the maximum achievable speedup over libjpeg-turbo's SIMD-version
can be written as 
\begin{equation}
	\mathit{Speedup} = \frac{\TTotal}{\THuffman},
\end{equation}
where $\TTotal$ is the decoding time of the SIMD-version. 
We compared the speedup of our approach to the theoretically attainable
 speedup in Figure~\ref{fig:pcnt_achvd}.  PPS stabilizes at an average speedup
of 88\% and attains its peak at 95\% of the theoretically attainable speedup.
For small images, the speedup was slightly higher than half of the maximum
attainable speedup because these images were partitioned into few chunks for
pipeline execution. 

Consequently, less work was executed in parallel with entropy decoding.
Increasing the number of chunks would result in a lack of data for GPU computation.  
As image size increases, an image is split into more chunks, and thus, less work of the parallelizable phase is visible to the user. 

\begin{figure}[htb]
    \begin{center}
        \includegraphics[clip=true]{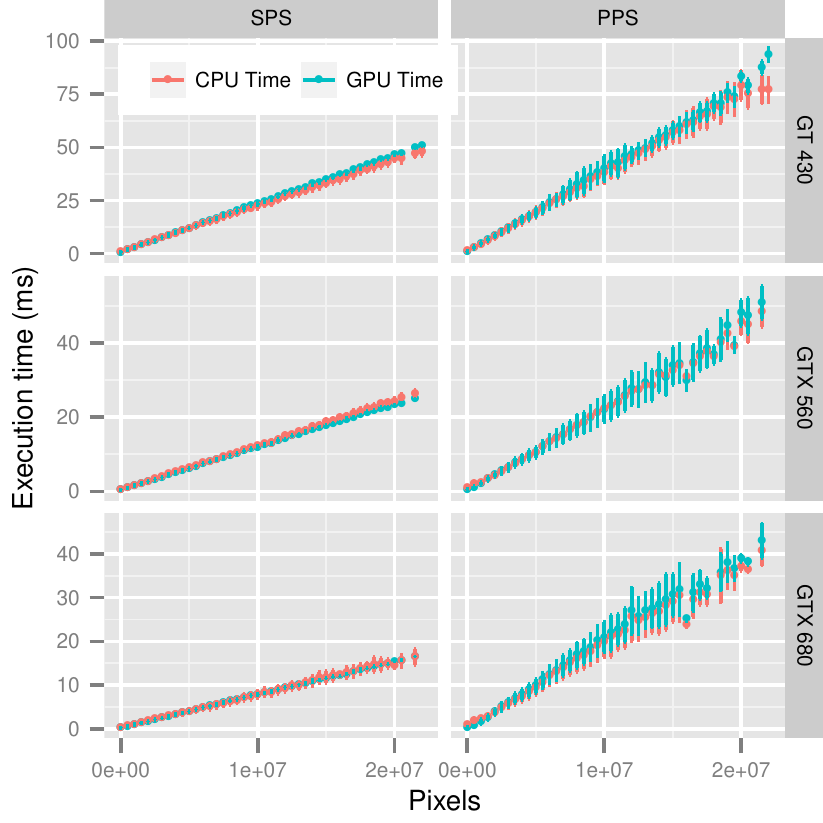}
    \caption{The average CPU and GPU execution time with standard deviation during parallel executions are balanced indicating balance workload between architectures.
    \label{fig:Prediction}}
    \end{center}
\end{figure}

Figure~\ref{fig:Prediction} shows execution times of the CPU and the GPU during
the parallel execution.  In the simple heterogeneous execution, the entropy
decoding time was omitted from CPU time as it is sequentially executed on the GPU.
Similarly, the entropy decoding of the first image chunk of the pipelined
execution was omitted.  GPU and CPU shared similar execution times indicating
well-balanced loads.  The main contribution to the variation in the
pipelined heterogeneous mode is Huffman decoding time.
%

\section{Related Work}
\label{sec:related}

DCT and IDCT algorithms are computationally intensive, but exhibit a high potential for parallelism. 
Various image processing algorithms, including DCT and IDCT, have been implemented for a GPU architecture by Yang et al.~\cite{yang2008parallel}. 
The authors utilized CUDA and applied CUFFT~\cite{cufftGuideV5}, a CUDA fast Fourier transform (FFT) library, to perform DCT and IDCT. 
However, extracting DCT from FFT introduces extra computational overhead.
The NVIDIA GPU Computing SDK provides DCT and IDCT sample code to demonstrate GPU programming. 
The kernel's input and output data type are float. 
In contrast to our implementation, the input buffer to our kernel is an array of short and the output buffer is an array of unsigned character. 
These data types are vectorized to minimize global memory transfer overhead.
Moreover, we combine dequantization, IDCT and color conversion in a single kernel to reduce data communication.

A task-parallel implementation of JPEG decoding using libjpeg-turbo has been explored by Hong et al.~\cite{hong2012task}.
Fork/join parallelism is applied to decode an image on CPU cores simultaneously. 
GPUJPEG~\cite{gpujpeg} is an open source JPEG image compression and decompression library for NVIDIA designed for real-time video.
Tasks for image decoding are divided between a CPU and GPUs where the CPU performs file I/O operations and Huffman decoding while the GPUs compute IDCT and color conversion.
The GPU kernel is, yet, optimized. 
Although the computationally intensive tasks are parallelized on the GPUs, the entire process is done serially.
On the contrary, we minimize hardware idle time by utilizing software pipelines and distributing workload across CPU and GPU.

Research on heterogeneous computing is receiving attention in high performance computing. 
Shee et al.~\cite{shee2008heterjpeg} conducted a case study on JPEG encoders on Application Specific Instruction-set Processors (ASIPs). 
They evaluated two parallel programming patterns: master-slave and pipeline. 
The master-slave model utilized task management and data-parallelism.
In the pipelined model, different ASIP processors were responsible for different stages of JPEG encoding. 
L. Chen et al.~\cite{MapReduceCPUGPU2012} proposed a similar idea for MapReduce applications. 
The authors developed two scheduling schemes, namely master-slave and pipeline model, on integrated CPU-GPU systems. 
Data-parallelism and pipeline-parallelism were utilized separately.
The Qilin framework~\cite{QilinFramework} and CHC framework~\cite{CHCFramework} showed possibilities of a cooperative CPU-GPU computation of a CUDA application. 
Both frameworks dynamically partitioned the workloads using their profiling based partitioning models. 
Qilin used an empirical approach recording new execution to a database while CHC applied a heuristic approach.
The partitioning schemes were designed for CUDA applications and only supported data-parallelism. 
The CAP scheduler~\cite{CAP2013} supports dynamic workload scheduling on CPU-GPU systems.
Profiling and workload partitioning are performed at run-time. 
CAP profiles a small portion of the workload, verifies the accuracy of the ratio and then uses the ratio for the remaining of workload. 
Although it can effectively partition workload, it only supports data-parallelism. 
Our proposed partitioning scheme, in comparison, is designed specifically for JPEG decoder. 
The workload is partitioned without user intervention, and the CPU and the GPU jointly perform the decoding tasks cooperating data-, task- and pipeline-parallelism.

\section{Conclusions}
\label{sec:conclusions}
We have introduced a novel JPEG decoding scheme for heterogeneous
architectures consisting of a CPU and an OpenCL-programmable GPU.  Our
method employs an
offline profiling step to determine the performance of a system's CPU and GPU
with respect to JPEG decoding.
We apply multivariate polynomial regression analysis 
to derive closed forms that characterize the performance of a given
CPU-GPU combination.  Image entropy and the image dimensions are 
the sole parameters for our performance model.  At run-time, closed forms
are evaluated for a given image to estimate execution times and load-balance
the decoding workload between the CPU and the GPU.
Our
run-time partitioning scheme exploits task, data and pipeline
parallelism by scheduling the non-parallelizable entropy decoding task on the
CPU, whereas IDCT, color conversion and
upsampling are conducted on both the CPU and the GPU.  Our kernels have been
optimized for GPU memory hierarchies.

We have implemented the proposed method in the context of the libjpeg-turbo
library, which is an industrial-strength JPEG encoding and decoding engine.
Irrespective of the GPU's computational power, our heterogeneous partitioning scheme
always achieves an improvement over the SIMD-version of libjpeg-turbo.  The
results show speedups up to 8.5x over the sequential version and up to 4.2x
over the SIMD version of libjpeg-turbo.  We have shown that our approach
achieves up to 95\% of the theoretically attainable speedup, with an average of 88\%.  With
the availability of GPU accelerators on desktops and embedded devices such as
tablets and smartphones, heterogeneous JPEG image decompression will enhance
image viewing experiences ranging from personal photos to very large image
applications in medical imaging and astronomy.
%
%
%


\bibliographystyle{abbrvnat}
\bibliography{libjpeg_pmam}
%
\end{document}